\documentclass[a4paper,11pt]{article}
\usepackage{graphicx,amsmath, amssymb}
\usepackage{lscape}
\usepackage{color, bm}
\usepackage{cite}
\usepackage{enumitem}

\newcommand{\preprintnumber}[1]{\vskip -2cm\vbox{  \baselineskip 14pt \hfill
    \hbox{\normalsize } \\
\hfill \hbox{\normalsize #1} } \vskip 2cm}
\newcommand{\email}[1]{\footnote{email:#1}}

\DeclareMathOperator{\tr}{tr}

\renewcommand{\O}{{\cal O}}
\renewcommand{\P}{{\mathbb P}}
\renewcommand{\r}{{r_{\rm CHC}}}
\newcommand{\V}{{\cal V}}
\newcommand{\Z}{{\mathbb Z}}
\newcommand{\rD}{\mathrm{D}}
	
\newcommand{\p}{\partial}

\newcommand{\be}{\begin{equation}}
\newcommand{\ee}{\end{equation}}

\begin{document}

\title{\preprintnumber{APCTP Pre2021-012, CTPU-PTC-21-27} 
Aligned Natural Inflation in the Large Volume Scenario}
\author{Stephen Angus$^{\rm a}$\email{stephen.angus@apctp.org}\ ,
Kang-Sin Choi$^{\rm b,c}$\email{kangsin@ewha.ac.kr}\ \
and Chang Sub Shin$^{\rm d}$\email{csshin@ibs.re.kr} \\
\it \normalsize $^{\rm a}$ Asia Pacific Center for Theoretical Physics, Postech, Pohang 37673, Korea \\
\it \normalsize $^{\rm b}$ Institute of Mathematical Sciences, Ewha Womans University, Seoul 03760, Korea \\
\it \normalsize $^{\rm c}$ Scranton Honors Program, Ewha Womans University, Seoul 03760, Korea\\
\it \normalsize $^{\rm d}$ Center for Theoretical Physics of the Universe, \\
\it \normalsize Institute for Basic Science, Daejeon 34126, Korea
}
\date{}

\maketitle

\begin{abstract}
We embed natural inflation in an explict string theory model and derive observables in cosmology. We achieve this by compactifying the type IIB string on a Calabi--Yau orientifold, stabilizing moduli via the Large Volume Scenario, and configuring axions using D7-brane stacks. In order to obtain a large effective decay constant, we employ the Kim--Nilles--Peloso alignment mechanism, with the required multiple axions arising naturally from anisotropic bulk geometries. The bulk volumes, and hence the axion decay constants, are stabilized by generalized one-loop corrections and subject to various conditions: the K\"ahler cone condition on the string geometry; the convex hull condition of the weak gravity conjecture; and the constraint from the power spectrum of scalar perturbations.  We find that all constraints can be satisfied in a geometry with relatively small volume and thus heavy bulk axion mass.  We also covariantize the convex hull condition for the axion-dilaton-instanton system and verify the normalization of the extremal bound.
\end{abstract}

\newpage

\section{Introduction}

It is of utmost importance to test string theory against observations.  Axions, or axion-like particles, are a ubiquitous feature of string compactifications:  
the size and shape of the extra-dimensional geometry are parametrized by scalar moduli fields, and under supersymmetry these moduli are naturally paired with axions.  They may arise from various antisymmetric tensor fields wrapping cycles in the compact dimensions.  Meanwhile, axions can play various roles in cosmology, resulting in phenomenological models that can be tested against data.  The physics of axions can therefore be used to constrain string theory, providing a direct link between real-world observations and the mathematical properties of the compact geometry. 

At tree level, the moduli appearing in compactifications of string theory from ten to four spacetime dimensions are massless scalar fields parametrizing flat directions in field space, which can generate fifth forces and vary the couplings between fundamental particles with zero energy cost.  To avoid this potentially apocalyptic scenario, these moduli must be stabilized at the minimum of a scalar potential --- this is typically achieved by including quantum corrections, giving moduli masses that depend on the details of the compactification.

We consider type IIB string theory compactified on a Calabi--Yau orientifold, in which moduli stabilization is well-understood.  Examples of scenarios in which all types of moduli can be stabilized include the KKLT scenario \cite{Kachru:2003aw} and the Large Volume Scenario (LVS) \cite{Balasubramanian:2005zx}.  In this work we focus on the latter scenario, for two principal reasons.  First of all, the volume $\V$ of the compact manifold is naturally stabilized at large values without tuning, which in turn allows parametric control of various perturbative expansions in powers of $1/\V$.  Secondly, it always includes at least one light axion associated to the moduli parametrizing the large volume.
The mass of this type of axion is predicted to be
\begin{equation} 
 m_a \sim M_{\rm Pl} e^{-k \V^{2/3}} \, ,
\end{equation}
where $M_{\rm Pl}\simeq 2.4 \times 10^{18}\text{ GeV}$ is the reduced Planck mass and $k$ is a model-dependent order-one constant. 
Since we consider a scenario in which the axions do not couple to QCD, strictly speaking these are axion-like particles, however in this work we will liberally use the term `axion' to describe all such light pseudoscalar fields.

Motivated by supersymmetric solutions to the electroweak hierarchy problem, which predict superpartners of observed particles to appear around the TeV scale, most scenarios tend to consider a hierarchically large volume $\V$. In such cases, this axion becomes almost massless and contributes to dark radiation \cite{Cicoli:2012aq,Higaki:2012ar}. Its abundance is reflected in the number of effective relativistic species observed at the time of the Cosmic Microwave Background (CMB). The most recent observational results from Planck 2018 give \cite{Aghanim:2018eyx}
\begin{equation} \label{DRbound}
  N_{\text{eff}} = 2.92^{+0.36}_{-0.37} \qquad \text{($95\%$ confidence level)}\, ,
\end{equation}
which is consistent with the Standard Model prediction, $ N_{\text{eff}} \simeq 3.046$, and can be accounted for entirely by the known three generations of neutrinos.\footnote{If we include the direct astrophysical measurement of the Hubble constant $H_0$ by Riess et al \cite{Riess:2018uxu} gives $N_{\text{eff}} = 3.27\pm 0.15$ ($68\%$ confidence level) \cite{Aghanim:2018eyx}, allowing a small excess.} 
The predictions of LVS amount to an excess of $\Delta N_{\text{eff}} \simeq \mathcal{O}(1)$, which is robust against modifications and loop corrections \cite{Angus:2013zfa,Angus:2014bia,Hebecker:2014gka}.  Thus, there appears to be little room for dark radiation.\footnote{However, this tension can be somewhat alleviated if the axion couples to visible-sector gauge bosons \cite{Hebecker:2014gka}.} 

In this paper, we consider an alternative scenario in which the volume $\V$ is not so large.  In this case, an axion can become an inflaton and realize natural inflation \cite{Freese:1990rb,Adams:1992bn}. The shift symmetry of the axion guarantees a flat potential, protected from quantum corrections which would ruin the observed approximate scale invariance of the power spectrum of scalar curvature perturbations. However, in order to explain the observed data, such as the spectral index of the power spectrum, the axion excursion, parametrized by its decay constant, needs to be larger than the Planck scale. This trans-Planckian behavior brings about many problems, such as accessibility, naturalness and consistency with the weak gravity conjecture \cite{Banks:2003sx,ArkaniHamed:2006dz}. 

In general compactifications, the internal geometry is not isotropic but rather can be described by multiple sub-cycles. Some of these are stabilized small at high energy but there still exist a number of large ``bulk'' cycles, whose moduli give the dominant contribution to the volume $\V$. Thus, naturally there should be as many light axions as there are large bulk cycles. It is possible for some of these axions to {\em align}, such that a linear combination of them may constitute an almost flat direction with a large effective decay constant.  This can generate an efffective trans-Planckian decay constant from individual decay constants below the Planck scale, potentially circumventing the problems of natural inflation.

In the Kim--Nilles--Peloso (KNP) mechanism \cite{Kim:2004rp}, two axions and their instanton charges (the coefficients of the axions in the potential) are aligned, such that their charge vectors are nearly coincident.  Rotating to the mass eigenbasis reveals that the orthogonal component becomes light, obtaining a large effective large decay constant
$$ 
 f_{\text{eff}} \sim f_\xi / Q_{\xi} \, .
$$
Here $f_\xi$ is the decay constant for the light axion obtained under the basis rotation, which is around the same size as those of the original axions.  Rather, the enhancement arises from the small effective charge of this axion, $Q_{\xi} \ll 1$.

The decay constant may be further enhanced if the axion potential is obtained by gaugino condensation \cite{Long:2014dta}. If the potential comes from a strongly-coupled $SU(N)$ gauge theory, the effective charge rescales as $Q_{\xi} \to Q_{\xi} / N$, yielding further enhancement,
\begin{equation} \label{Nenhancement}
 f_{\text{eff}} \sim N f_\xi / Q_{\xi} \, .
\end{equation}
This $N$ is also related to the number of degenerate branches of vacua, corresponding to discrete phases of the gaugino condensate.  In arranging this we should ensure that there is no tunneling to other branches during inflation, which could cancel the new factor $N$. We will estimate the tunneling amplitude, which setting an upper limit on $N$.

Another approach is to consider alignment arising from a non-diagonal K\"ahler metric for multiple axions \cite{Bachlechner:2014gfa,Bachlechner:2015qja}.
This possibility has led to developments in multi-axion cosmology \cite{Choi:2014rja,Higaki:2014pja,Kappl:2014lra,Higaki:2014mwa,Kappl:2015esy,Choi:2015aem,Choi:2020rgn}.  String embeddings of this scenario were first studied in \cite{Long:2014dta} for the type IIB string and subsequently expanded to the heterotic string \cite{Ali:2014mra}. 

Axion alignment restores a (discrete) shift symmetry. It is believed that global symmetries should be broken by quantum gravity effects, thus constraining the possibility of alignment. This condition on global symmetries is quantified in the weak gravity conjecture \cite{ArkaniHamed:2006dz,Heidenreich:2015nta,Rudelius:2015xta}. This conjecture states that the charge measured in unit of the inverse decay constant should be larger than the Planck mass. Conversely, for a fixed charge, it restricts the upper limit of the allowed decay constant. For multiple axions, a stronger condition is required: the convex hull spanned by the charge-to-mass ratio vectors should be large enough to contain those of appropriate extremal states \cite{Cheung:2014vva,Rudelius:2015xta,Montero:2015ofa,Brown:2015lia,Brown:2015iha}.

It is interesting to investigate whether aligned natural inflation can be realized explicitly in string theory for the following reasons.  First of all, string theory is regarded as a consistent theory of quantum gravity, and thus it is expected that a string construction should automatically satisfy the weak gravity conjecture. Second, the parameters are in principle calculable, giving cosmological observables in terms of properties of the compact geometry.  Moreover, from a top-down perspective the string geometry itself can impose consistency constraints: for instance, the K\"ahler cone condition restricts the cycle volumes and hence the allowed decay constants.


We also refine the alignment mechanism and the weak gravity conjecture, by generalizing the axion decay constant to be matrix-valued (see also e.g. \cite{Bachlechner:2014hsa}) and carefully distinguishing between the decay constants and the instanton charges. We thus restate the KNP alignment scenario as a condition purely on the instanton charges. With this, we may take into account the enhancement from the charge rescaling (\ref{Nenhancement}).  Furthermore, precisely differentiating between the charges and the decay constants also allows us to make a more precise statement of the weak gravity conjecture for the axion-instanton system.  We verify, both from the perspective of the string embedding as well as consideration of extremal wormholes, that the normalization of the axionic weak gravity conjecture should be $\sqrt{2/3}$. 

The paper is organized as follows. In Section 2 we review aligned natural inflation 
and consider possible restrictions from tunneling. We also clarify the meaning of the decay constant and the instanton charges, which are required for clarifying the weak gravity conjecture. In Section 3, we prepare the ingredients of our construction, first reviewing relevant components of four-dimensional reductions of type IIB supergravity reductions, writing down a string-originated effective Lagrangian describing axions, and discussing choices of geometry that lead to multi-axion potentials.  We identify axion charges and decay constants as well as instanton actions from the supergravity viewpoint.
We use this to refine the axion-instanton form of the weak gravity conjecture, along with its multi-axion generalization. 
This setup turns out to be sufficient to pinpoint the extremal bound and fix the convex hull condition precisely.

In Section 4, we present our explicit embedding of aligned natural inflation into a concrete type IIB string theory model.  After discussing moduli stabilization in the Large Volume Scenario, we derive observable quantities. We look for benchmark points for the axion decay constants and consider the possibility of alignment. The convex hull condition of the weak gravity conjecture and the range of the K\"ahler parameters show preferences in the opposite directions, setting bounds on the allowed ratios of the moduli. Then we compare our predictions with observations in the Planck 2018 results. We discuss some interesting findings and future directions in the final section.

\section{Axions and aligned natural inflation} \label{sec:aligned}

In this section we review natural inflation and the alignment mechanism from multiple axions, which allows for sufficiently small slow-roll and curvature parameters without requiring field excursions greater than the Planck length. We also briefly explore a further possible enhancement mechanism arising from gaugino condensation as well as its stability against vacuum tunneling.

\subsection{Natural inflation}
An axion-like particle $a$ can undergo natural inflation \cite{Freese:1990rb,Adams:1992bn}. This scenario solves $\eta$-problem of inflation since the axion is a pseudo-Goldstone boson associated with a shift symmetry, and thus the potential is protected from radiative corrections.  The axion potential is generated only by non-perturbative effects,
\begin{equation}
 V = \Lambda^4 \left(1- \cos \frac{a}{f}\right) \, , \label{Vnatlinfl}
\end{equation}
with decay constant $f$ measured in units of the reduced Planck mass, $M_{\rm Pl}$.
The axion may roll down the potential, with slow-roll dynamics characterized by the parameters \cite{Kappl:2014lra}
\begin{align}
 \epsilon & \equiv \frac{M_{\rm Pl}^2}{2} \left(\frac{V'(x)}{V(x)}\right)^2 \, ,   \\
 \eta & \equiv M_{\rm Pl}^2 \frac{V''(x)}{V(x)} \, .
\end{align}

Observables relevant to inflation, of which we quote the Planck 2018 results in the $\Lambda$CDM model (Planck+lensing+BAO) \cite{Aghanim:2018eyx}, include the spectral index  (68\% confidence level),
\begin{equation} \label{specindex}
 n_s \simeq  1 - 6 \epsilon + 2 \eta  = 1 - f^{-2} - 4 \epsilon = 0.9665 \pm 0.0038 \, ,
\end{equation}
and the tensor-to-scalar ratio,
\begin{equation} \label{tensortoscalar}
r_{0.002} \simeq 16 \epsilon = 4 \left (1-n_s -  f^{-2} \right)< 0.106 \, .
\end{equation}

The spectral index (\ref{specindex}) is relatively robust, and obtaining sufficiently large $n_s$ in natural inflation requires a large enough decay constant,
\begin{equation} \label{fnatural}
 f \gtrsim 4 M_{\rm Pl} \, .
\end{equation}
An important question is whether we can obtain such a large $f$ \cite{Banks:2003sx} in a controlled model.  Since quantum corrections in string theory are suppressed only in a parameter regime below the Planck mass, this presents a challenge for string-theory model building.  One candidate resolution is to generate effectively super-Planckian decay constants using only sub-Planckian parameters via an alignment mechanism, which we introduce in the following subsection.

Another useful fact is that the power spectrum  of scalar perturbations ${\cal P}_\zeta$ determines the height of the inflationary potential \cite{Di},
\begin{equation} \label{heightpotential} 
 \frac{\Lambda^4}{M_{\rm Pl}^4} \simeq \frac{12 \pi^2 }{(2q^2+1)e^{N_e/f^2}} {\cal P}_\zeta \, ,
\end{equation}
where $N_e$ is the $e$-folding number. The latest Planck measurement of the scalar power spectrum amplitude gives ${\cal P}_\zeta \simeq 2.1 \times 10^{-10}$ \cite{Aghanim:2018eyx}.
Assuming $50 < N_e < 60$, for a  wide range of decay constants, $3 < f < 10$, this translates into $\Lambda^4 \simeq 10^{-9} M_{\rm Pl}^4,$ corresponding to an inflation scale of $\Lambda \simeq 10^{16} \rm ~ GeV$.

\subsection{Alignment of axions} \label{sec:alignment}

We now review the axion alignment mechanism from a bottom-up field theoretical perspective.  Later we will consider an explicit embedding into string theory, which will provide a more fundamental justification of this structure as well as an interpretation of the various parameters in terms of the underlying geometry.  Consistency conditions such as supersymmetry will provide further constraints. 

When multiple axions are present, a linear combination of them may obtain an effective decay constant which is much larger than those of the individual axions.  Consider the case of an anomalous symmetry containing Abelian factors $U(1)_j, j=1,\dots ,N$, as well non-Abelian gauge groups $G_I,I=1,\dots,M$, giving rise to a Lagrangian
 \begin{equation} \label{startingL}
 {\cal L} =  k_{ij} \partial_\mu \vartheta_i \partial^\mu \vartheta_j - \sum_{I=1}^{M} \Lambda_I^4  \left[1-\cos \left(Q_{Ij} \vartheta_j\right)\right] \,,
\end{equation}
where $k_{ij}$ is a real, symmetric K\"ahler metric and $\Lambda_I$ are coefficients depending on the theory in question. Here $Q_{Ij}$ are the coefficients of the $G_I$-$G_I$-$U(1)_j$ anomaly,
\begin{equation} \label{anomcoeff} 
 \partial_\mu J_j^5 = \frac{Q_{Ij}}{32\pi^2} \tr \epsilon^{\mu \nu \rho \sigma} F_{\mu \nu,I} F_{\rho \sigma,I} \, , \qquad Q_{Ij} = \tr  t_I^2 q_j \, ,
\end{equation}
which are integrally quantized with normalization $\tr t^a t^b = \delta^{ab}$, in the case of the fundamental representation of $SU(N)$.  It follows that the axion fields have well-defined periodicity, $\vartheta_i \to \vartheta_i + 2\pi$. 

For now and in the remainder of this discussion, we focus on the minimal configuration that captures the essential physics, which is the case of two instanton contributions. With a diagonal K\"ahler metric, $k_{i j}=\text{diag}(f_1^2,f_2^2)/2$, the canonically normalized fields are
\begin{equation} \label{CanonicalAxions}
  \phi_1= f_1\vartheta_1^{\text{diag}} \, , \qquad {\phi_2} = f_2\vartheta_2^{\text{diag}} \, .
\end{equation}
  The Lagrangian (\ref{startingL}) provides a  lattice basis, which can be expanded as
\begin{equation}
 {\cal L} = \frac12 \partial_\mu \phi_1 \partial^\mu \phi_1 + \frac12 \partial_\mu \phi_2 \partial^\mu \phi_2  -V \, ,
\end{equation}
where
 \begin{equation} \label{axionLattice}
V= \Lambda_1^4 \left[1-  \cos \left( \frac{Q_{11}}{f_1} \phi_1 + \frac{Q_{12}}{f_2}\phi_2 \right)\right] -\Lambda_2^4 \left[1-  \cos \left(\frac{Q_{21}}{f_1} \phi_1 +\frac{ Q_{22}}{f_2} \phi_2\right)\right] \, .
\end{equation}

In the alignment limit \cite{Kim:2004rp},
\begin{equation} \label{alignment}
 Q_{11}: Q_{12} = Q_{21}: Q_{22} \, ,
\end{equation}
the two axions become coincident because the arguments of the two potentials in (\ref{axionLattice}) become proportional. Since the total degrees of freedom should be preserved, there should be an orthogonal axion whose potential is flat. This is the basis of the Kim--Nilles--Peloso (KNP) mechanism, which makes use of the fact that the orthogonal axion direction obtains an effectively large decay constant \cite{Kim:2004rp}.

In this paper we take some care to separate axion charges and decay constants, such that we can define alignment purely in terms of charges.
To see how the alignment manifests, redefine the basis by an $SO(2)$ rotation 
\begin{equation} \label{alignedaxions}
 \left(\begin{array}{c} \phi_\xi \\ \psi_\psi\end{array}\right) = \frac{1}{\sqrt{f_1^2 Q_{12}^2 + f_2^2 Q_{11}^2}}
 \left(\begin{array}{cc} f_1 Q_{12} & - f_2 Q_{11} \\ f_2 Q_{11} & f_1 Q_{12} \end{array}\right)\left(\begin{array}{c} \phi_1 \\ \phi_2\end{array}\right) \, .
\end{equation} 
{The first term of the potential (\ref{axionLattice}) then becomes} a function of $\phi_\xi$ only.  
For each of $\phi_\xi$ and $\phi_\psi$, we can define the respective decay constants as
\begin{equation} \label{fxi}  f_\psi = \frac{f_1 f_2 \sqrt{Q_{11}^2+ Q_{12}^2}}{\sqrt{f_1^2 Q_{12}^2 + f_2^2 Q_{11}^2}}\, , \qquad f_\xi = \frac{\sqrt{f_1^2 Q_{12}^2+ f_2^2 Q_{11}^2}}{\sqrt{Q_{11}^2+ Q_{12}^2}}\, .
\end{equation} 
Note that there is no special hierarchy between $\{f_\psi, f_\xi\}$ and $\{f_1, f_2\}$. In this basis, the potential can be expressed as
\begin{equation} \label{alignedpotential} \begin{split}
 V =   \Lambda_1^4 \left[1-  \cos \left( \frac{Q_{1 \psi}}{f_\psi}\phi_\psi \right)\right] + \Lambda_2^4 \left[1-  \cos \left(\frac{Q_{\xi}}{f_\xi} \phi_\xi +\frac{ Q_{2\psi}}{f_\psi} \phi_\psi\right)\right]\end{split} \, , \end{equation}
with the charges
\begin{equation} \label{effcharge}\begin{split}
Q_{1\psi} &= \sqrt{Q_{11}^2+ Q_{12}^2}\, ,\qquad Q_{\xi}  = \displaystyle \frac{ Q_{21} Q_{12}-Q_{11} Q_{22}}{ \sqrt{Q_{11}^2+ Q_{12}^2}} \, , \\
Q_{2\psi} &= \left( \frac{f_1^2 Q_{12} Q_{22}+f_2^2 Q_{11} Q_{21} }{f_1^2 Q_{12}^2 + f_2^2 Q_{11}^2 }\right) Q_{1\psi} \, .
\end{split}\end{equation}
Thus, the alignment limit (\ref{alignment}) can be achieved by choosing $Q_{Ii}$ to satisfy  
\begin{equation} \label{misalignment}
|Q_\xi| \ll 1 \, . 
\end{equation}
As a result, the effective decay constant of $\phi_\xi$, $f_{\rm eff}$, from the second term of the potential,
becomes hierarchically larger than $f_\xi$ as 
\begin{equation}  \label{feff}
 f_{\rm eff} = \frac{ f_{\xi}}{ |Q_{\xi}|} =  \frac{\sqrt{f_1^2 Q_{12}^2 + f_2^2 Q_{11}^2}}{|Q_{21} Q_{12} - Q_{11} Q_{22}|} \, .
\end{equation}

One can simplify the dynamics of the light axion by integrating out the heavy degree of freedom. 
Depending on the origin of the terms in the potential, there can be a natural hierarchy between $\Lambda_1^4$ and $\Lambda_2^4$. 
Without loss of generality, by assuming $\Lambda_1^4 \gg \Lambda_2^4$ we can easily integrate out the heavy axion ($a_H\simeq \phi_\psi$) from the first term of the potential,
$$
 \left \langle \frac{Q_{1\psi} \phi_\psi}{f_\psi} \right \rangle  = 2\pi k 
  + {\cal O}\left(\frac{\Lambda_2^4Q_{2\psi}^2}{\Lambda_1^4 Q_{1\psi}^2}\right) \quad {\rm for} \ 
  k \in \mathbb{Z} \, .
$$
The resulting effective Lagrangian for the light axion ($a \simeq \phi_\xi$)  becomes
\begin{equation}\begin{split} \label{effLagrangian}
{\cal L}_{\rm eff} &= \frac{1}{2} (\partial_\mu a)^2 - \Lambda_2^4 \left[1- \cos \left(\frac{a}{f_{\rm eff}} + \frac{Q_{2\psi}}{Q_{1\psi}}2\pi k \right)\right] \\
& \quad +  \frac{\Lambda_2^8 Q_{2\psi}^2}{2\Lambda_1^4 Q_{1\psi}^2} \sin^2\left(\frac{a}{f_{\rm eff}} + \frac{Q_{2\psi}}{Q_{1\psi}} 2\pi k\right) + {\cal O}\left(\frac{\Lambda_2^{12}Q_{2\psi}^4}{\Lambda_1^8 Q_{1\psi}^4}\right) \, .
\end{split}
\end{equation}
The leading term of the potential is simplified to the form of (\ref{Vnatlinfl}), with decay constant $f_{\rm eff} \gg f_{1,2}$. In the practical analysis to obtain cosmological parameters, we take into account the dynamics of both axions without approximations. Note that the existence of such a light direction does not rely on a hierarchy between $\Lambda_1^4$ and $\Lambda_2^4$.

For a non-diagonal K\"ahler metric, we can follow essentially the same procedure. 
Since $k_{ij}$ is a real, symmetric matrix, we may diagonalize it via an orthogonal matrix $P^{-1} = P^{\top}$ such that
$$
P_{ik}^{\top}(k_{kl}) P_{lj} = \frac12 \begin{pmatrix} f_{1}^2 & 0 \\ 0 & f_{2}^2 \end{pmatrix}\, , \qquad \vartheta^{\text{diag}}_i = P^{\top}_{ij} \vartheta_j \, ,
$$
where the eigenvalues are positive semi-definite.
After canonically normalizing as in (\ref{CanonicalAxions}), we obtain
\begin{equation} \label{axionpotential}
\begin{split} 
\ {\cal L} &= \frac12 \partial_\mu \phi_1 \partial^\mu \phi_1 + \frac12 \partial_\mu \phi_2 \partial^\mu \phi_2 \\
&\quad -\sum_{I=1,2} \Lambda_I^4  \left[1- \cos \left( \frac{Q_{I k} P_{k1}}{f_1} \phi_1 +  \frac{ Q_{I k} P_{k2}}{f_2}\phi_2 \right)\right]\,.
 \end{split} \end{equation}
 Redefining 
 $Q_{Ik} P_{kj} \equiv Q'_{Ij} $, the Lagrangian formally reduces to (\ref{axionLattice}). 
Obviously, for a general K\"ahler metric the potential is not necessarily periodic with $\phi_i \to \phi_i + 2\pi f_i$ because the elements $P_{kj}$ may not be integers. Rather, in this expression the periodicity of $\vartheta_i$ is hidden. Nevertheless, each quantity $f_i/Q_{Ii}'$ still plays the same role as $f_i/Q_{Ii}$ in (\ref{axionLattice}) when we consider the dynamics of the axions. 
Therefore,  the alignment condition for the primed charges can be discussed in the same way. In fact, for any $P_{kj}$, the alignment condition reduces to that of the unprimed charges $Q_{Ik}$. More explicitly, from (\ref{effcharge}) we find that the effective charge $Q_\xi'$ is 
$$
Q_\xi' =  \frac{Q_{21}' Q_{12}' - Q_{11}' Q_{22}'}{\sqrt{Q_{11}'^2 + Q_{12}'^2}} = Q_\xi \, .
$$
The alignment limit $|Q_\xi|\ll 1$ is thus equivalent to $|Q_\xi'|\ll 1$.

Alternatively, we may define the matrix-valued decay constant ${\bf f}$ as one satisfying
\begin{equation} \label{decayconstmatrixnorm}
2k_{ij} =  ({\bf f}^\top{\bf f})_{ij} =  {\bf f}_{ki} {\bf f}_{kj} \, . 
\end{equation} 
It is known that we may always find a symmetric matrix ${\bf f}$ with this property \cite{Bachlechner:2014hsa}.
Then the canonically normalized axion is defined as
\begin{equation} \label{axiondefinition}
 \phi_i \equiv  {\bf f}_{ij} \vartheta_j \, .
\end{equation}
The matrix-valued decay constant ${\bf f}$ is related to $P$ as ${\bf f}_{ij} = f_i P^\top_{ij}$ (no sum over $i$). The potential can be written as
\begin{equation} 
\begin{split}
V&= \sum_{I=1,2} \Lambda_I^4  \Big[1- \cos \left(   Q_{Ik}{\bf f}^{-1}_{k1}\phi_1 +  Q_{Ik} {\bf f}^{-1}_{k2}\phi_2 \right)\Big] \, .
 \end{split} \end{equation}
 Note that we have again separated the charges and the decay constants. Still in this case, the alignment limit can be expressed purely in terms of the charges as in (\ref{alignment}). This is essentially because the decay constant is well-defined as dependent on the axion flavor $i$ as in (\ref{axiondefinition}), not the instanton flavor $I$. We stress that each of them rotates covariantly under a basis transformation and the size of the decay constants remain of the same order, verifying that the alignment arises from the special combination of {\em charges}. Following the same logic as used to obtain (\ref{effLagrangian}), 
the general expression for the effective decay rate of the light axion $\phi_\xi$ becomes  
 \begin{equation}\label{feffgen}
 f_{\xi} = \frac{ \sqrt{ ({\bf f}_{ij}\epsilon_{jk} Q_{1k})  ({\bf f}_{ij'}\epsilon_{j'k'} Q_{1k'})} }{\sqrt{Q_{11}^2+ Q_{12}^2}} \, , \qquad 
 f_{\rm eff} = \frac{f_\xi}{Q_\xi} \, , \end{equation}
where $\epsilon_{12}=-\epsilon_{21}=1$.

Finally, there is a room for further improvement by utilizing off-diagonal terms of the K\"ahler metric. In the two-axion case, if the rotation matrix becomes maximal, 
\begin{equation} \label{MaxRotMatrix}
  P = \frac{1}{\sqrt{2}}\begin{pmatrix} 1 & 1 \\ -1 & 1 \end{pmatrix}\, .
\end{equation}
then (\ref{feffgen}) becomes 
\begin{equation}\label{feffkinetic}
f_{\rm eff} = \frac{\sqrt{f_1^2\left(\frac{Q_{11} + Q_{12}}{\sqrt{2}}\right)^2 + f_2^2\left(\frac{Q_{11} - Q_{12}}{\sqrt{2}}\right)^2}}{|Q_{21} Q_{12} - Q_{11} Q_{22}|} \, .
\end{equation}
Comparing (\ref{feffkinetic}) with the simple diagonal example (\ref{feff}), we see that 
for $f_1= f_2$, the two expressions are the same regardless of the detailed charge assignment. However, if there is some hierarchy between the eigenvalues of ${\bf f}$, such as $f_1 \gg f_2$,  
a clever charge assignment can provide an additional enhancement factor. For instance, 
if $Q_{11} = Q_{12}$, 
(\ref{feff}) becomes $f_{\rm eff}\simeq f_1/|Q_{21}- Q_{22}|$, while (\ref{feffkinetic}) 
becomes $f_{\rm eff} \simeq \sqrt{2}f_1/|Q_{21}-Q_{22}|$. This is an example of
kinetic alignment \cite{Bachlechner:2014hsa,Higaki:2014mwa}, in which 
 alignment between the kinetic metric and the charges is imposed.

For $N$ axions, except for charge alignment scenarios like the clockwork mechanism \cite{Choi:2014rja,Choi:2015fiu,Kaplan:2015fuy}, we may also obtain an enhancement factor $\sqrt{N}$ simply from the kinetic metric, due to the number of summed terms appearing inside the square root in (\ref{feffgen}), as in the N-flation scenario. 
Strategic alignment between the kinetic metric and the charges can also provide additional enhancement if the eigenvalues of the decay constant have a hierarchical structure. The distribution of enhancement factors for randomly generated K\"ahler metrics, as well as the possibility of both kinds of enhancement, was studied in Ref. \cite{Bachlechner:2014gfa}.


\subsection{Further enhancement and tunneling effects} \label{sec:tunneling}

There is further room for enhancement of the axion decay constant. 
If the axion is obtained from gaugino condensation in a supersymmetric gauge theory, as will be discussed in detail in Section \ref{sec:axionorigin},
the decay constant may increase by a factor of $N$, the rank of the $U(N)$ gauge group.

In gaugino condensation of a strongly coupled $U(N)$ gauge theory, the axion arises as the imaginary part of a complex field $T$, $\vartheta={\rm Im}T$, appearing in the gaugino condensate superpotential and its resulting scalar potential,
$$
 W \sim e^{-T/N} \to V \sim \cos \frac{\vartheta}{N} = \cos \frac {a}{N f} \, . 
$$
It appears that this would lead to enhancement of the decay constant $f$ by a factor of $N$. 
However, it is not immediately obvious that this would happen in practice, since there may be degenerate vacua which may lead to tunneling between the different branches of the axion potential. 
Similarly, in the presence of multiple axions, any tunneling between different minima would ruin the separation between the axions, in turn ruining the alignment. 
This presents a general problem for the alignment scenario.  Here we analyze this possibility by computing the tunneling rate between adjacent branches of the axion potential.

As a warmup, we first analyze the enhancement using a simplified aligned two-axion model, which captures the essential physics of tunneling between different branches. We take a Lagrangian of the form
\be
{\cal L} = \frac{1}{2}  f_1^2(\partial_\mu \vartheta_1)^2 + \frac{1}{2}  f_2^2 (\partial_\mu \vartheta_2)^2 + \Lambda_H^4 \cos (N \vartheta_1 - \vartheta_2) + \Lambda_L^4 \cos(\vartheta_1) \, ,
\ee  
where the index ``$H$" implies that the corresponding potential is the dominant source making one of the axions heavy, while ``$L$" corresponds to the potential that is relevant for the dynamics of the light axion.
In the alignment limit, integrating out the heavy axion ($a_H\simeq f_1\vartheta_1$) from the first potential term  induces the following effective Lagrangian of the light axion:
\be \begin{split}\label{Vaeff}
{\cal L}_{\rm eff} &= \frac{1}{2} \left(f_2^2 + \frac{f_1^2}{N^2}\right) (\partial_\mu\vartheta_2)^2 + \Lambda_L^4\cos\left(\frac{\vartheta_2}{N} + \frac{2\pi k}{N}\right) \\ 
&= \frac{1}{2} (\partial_\mu a)^2 + \Lambda_L^4 \cos\left(\frac{a}{N f_a} + \frac{2\pi k}{N}\right) \quad {\rm for}\ k \in \mathbb{Z}\,.
\end{split} \ee 
Canonical normalization determines the light axion, $a \equiv f_a \vartheta_2$, with 
\be 
 f_{\rm eff} = Nf_a= \sqrt{f_1^2 +  N^2 f_2^2}\,.
\ee  
Here $f_{\rm eff}$ is enhanced by a factor $N$ compared to $f_a$, while there are $N$ branches labeled by $k=0,\cdots, N-1$. For a given branch (e.g. $k=0$), during an excursion of the axion in the range $0 \le a \le \pi f_{\rm eff}$, quantum tunneling may allow a rapid transition to other branches with lower values of the potential.
In such a case, the axion, as the inflaton, can only travel slowly for a distance of ${\cal O}(f_a)$ due to the deformed scalar potential, cancelling the enhancement factor $N$. 

To obtain the tunneling rate between different branches (for instance, from $k=0$ to $-1$), we consider the nucleation rate for the  critical bubble which connects two local minima along the direction of the heavy axion with the field displacement $\Delta a_H\simeq f_1\Delta \vartheta_1= 2\pi f_1/N$.  Tunneling can take place whenever the potential at $k=-1$ is lower than the potential at $k=0$. 
The potential difference between the two positions is given as 
\begin{equation} \label{potdiff}
	\Delta V_L\equiv V_L(k=0) - V_L(k=-1) > 0 \, ,
\end{equation}
where $V_L$ is the light axion potential from (\ref{Vaeff}). 
The height of the potential barrier between the two local minima along the heavy axion direction is of ${\cal O}(\Lambda_H^4)$, and it is hierarchically larger than $\Delta V_L$, i.e. $\Lambda_H^4 \gg \Lambda_L^4\gtrsim \Delta V_L$. Therefore, one can use the thin wall approximation to evaluate the bounce solution following \cite{Coleman:1977py}. Ignoring gravitational effects, the transition rate is given by 
\begin{equation} 
\Gamma_{\rm tunnel} \propto e^{ - S_E} \, ,
\end{equation} 
where $S_E$ is the bounce action
\begin{equation} 
S_E = \frac{27\pi^2}{2} \frac{\sigma_B^4}{(\Delta V_L)^3} \, .
\end{equation} 
Here $\sigma_B$ is the tension of the bubble wall, to be calculated shortly, and $\Delta V_L$ is the potential difference between the branches (\ref{potdiff}).  

In principle,  since the potential is not a local minimum for the light axion either before or after tunneling, 
we should also consider the kinetic energy of $a$ and the field displacement $\Delta a$ during tunneling. 
However, owing to the fact that the light axion is slowly rolling during inflation, we can safely ignore the effect of the axion kinetic energy. 
Furthermore, the possibility of nonzero displacement $\Delta a$ of the light axion leads to a larger total field distance $\sqrt{\Delta a_H^2 + \Delta a^2}$ traversed during the tunneling process.  
As $\Delta a$ increases, the bubble wall tension increases more significantly compared to $\Delta V_L$, so that 
the corresponding bounce action also increases. Therefore, in order to obtain a conservative bound on the tunneling rate and observe its parametric dependence,  it is sufficient to consider the heavy-axion field direction only.
For more concrete discussion about the possible effect of the light field dynamics, see \cite{Brown:2016nqt}. We then estimate the tension as $\sigma_{\rm B} \simeq \Lambda_H^2 \Delta a_H$, giving
\be 
\Delta V_L\simeq \frac{2\pi }{N} \Lambda_L^4 \sin(a_{\rm inf}/f_{\rm eff})\, , \qquad
\sigma_{\rm B} = \frac{  c_B\pi }{N} \Lambda_H^2f_1 \, , 
\ee 
where $a_{\rm inf}$ is the axion value during inflation, and $c_B$ is an ${\cal O}(1)$ number. 
Therefore, the bounce action is
\be 
S_E \simeq \frac{27 c_B^4\pi^3}{16\sin(a_{\rm inf}/f_{\rm eff})^3 N} \left(\frac{f_1}{\Lambda_L}\right)^4 \left(\frac{\Lambda_H}{\Lambda_L}\right)^8 \, . \label{SEtoy}
\ee 
It is not difficult to obtain $S_E \geq {\cal O}(100)$ either from the hierarchy between $f_1$ and $\Lambda_L$ or between $\Lambda_H$ and $\Lambda_L$, thus tunneling is suppressed and the decay constant may be enhanced, $f_{\rm eff} = Nf_a$.  Since $N$ appears in the denominator, in effect this restricts $N$ to be small. To estimate the allowed value, we consider conservative values: $\sin(a_{\rm inf}/f_{\rm eff}) \simeq 1$, $c_B^4  \simeq 0.01$, $\Lambda_H/\Lambda_L \simeq 10$, $f_1/\Lambda \simeq 0.1$, giving $
S_E \simeq 5232/N$.
This shows that typical values, $N \lesssim 50$, are robust.

We can use the same analysis on the multi-axion model discussed in the previous section, where we allowed general instanton charges $Q_{Ij}$ as the coefficients of the respective axions $\vartheta_j$.  Again we focus on the two-axion case.  In the alignment limit, the heavy axion $\phi_\psi$ may be integrated out, leading to the possibility of degenerate vacua depending on the ratio of charges.  Then although we have a light axion with a large effective decay constant, again there can potentially be tunneling between different minima, ruining the alignment.

	
Before integrating out heavy axion $\phi_\psi$, the Lagrangian (\ref{alignedpotential}) shows that its excursion distance is $\Delta \phi_\psi \simeq 2\pi f_\psi /Q_{1\psi},$ giving the bubble tension
$$
\sigma_{B} = \frac{c_B\pi}{\sqrt{Q_{11}^2 + Q_{12}^2}}\Lambda_1^2 f_\psi.
$$
Note that in the alignment limit, $Q_{2\psi}/Q_{1\psi} \to Q_{21}/Q_{11}$.
The local minima of the first potential gives $\langle Q_{1\psi}\phi_\psi/f_\psi\rangle = 2\pi k$, which  leads to $Q_{11}/{\rm gcd}(Q_{11},Q_{21}) \equiv N'$ number of branches for the light axion. In most cases, $Q_{11},Q_{21}$ are unrelated and they can be easily relatively prime so that $N' = Q_{11}.$ In the worst case, we have $N'=2$. If $1/N'$ is an integer, we do not have degenerate vacua.
For the effective Lagrangian (\ref{effLagrangian}), 
the transition between the branches $k=0$ and $k=-1$ gives 
\begin{equation} 
\Delta V_L  \simeq\frac{2\pi }{N'} \Lambda_2^4 \sin\frac{a_{\rm inf}}{f_{\rm eff}},\quad 
\end{equation} 
Then, the bounce action is 
\begin{equation} \label{bounce}
S_E \simeq \frac{27c_B^4\pi^3 N^{\prime 3}}{16 \sin (a_{\rm inf}/f_{\rm eff})^3(Q_{11}^2+ Q_{12}^2)^2} 
\left(\frac{ f_\psi}{\Lambda_2}\right)^4   \left(\frac{\Lambda_1}{\Lambda_2}\right)^8.
\end{equation}

This time, the tunneling may prohibit large charges $Q_{11},Q_{12}$. Assuming $Q_{11} \simeq Q_{12} \equiv Q $, we estimate similarly as above $S_E \simeq  1308N^{\prime 3}/Q^4$, requiring $Q_{Ij} \lesssim 3N^{\prime 3/4}$.

Finally, we may combine both effects. The restriction on the rank of the gauge group from the initial toy model can be directly applied to the aligned axion potential. As a result, each instanton charge $Q_{Ij}$ is effectively rescaled by the rank $N_I$ of the corresponding gauge group,
\begin{equation} \label{instchargesuppr}
	Q_{Ij} \to Q_{Ij} / N_I \, .
\end{equation}
Then the bounce action is rescaled as $S \to S N_1^4 $, 
which allows larger charges $Q_{Ii}$ due to the effective suppression by $N_I$ in (\ref{instchargesuppr}), giving the less stringent bound
\begin{equation} \label{tunnelingaligned}
 Q_{Ij} \lesssim 3N_I N^{\prime 3/4} \, .
\end{equation}

\section{Natural inflation from string theory}

We now embed the above axion alignment scenario into type IIB string theory.
In preparation for our explicit model which we present in section \ref{sec:explicitmodel}, we first review how axions arise in string compactifications and explore how the decay constants are determined by the underlying geometry via moduli.  Furthermore, the axion-instanton system is subject to the weak gravity conjecture, the implications of which we examine in detail. 

\subsection{String origin of axions} \label{sec:axionorigin}

Consider type IIB string theory compactified on a Calabi--Yau orientifold $X$.  
At energies below the compactification scale, the physics should be described by an effective four-dimensional $\mathcal{N} = 1$ supergravity theory with various massless scalar fields: the axio-dilaton $S = e^{-\phi} + iC_{0}$, where $g_{s} = e^{\phi}$ gives the string coupling, $C_0$ is the RR scalar, K\"{a}hler moduli $T_{i} = \tau_{i} + ib_i$  that we define below, and complex structure moduli $U_{a} = u_{a} + ic_{a}$ associated with the shape of the internal manifold.  Here and in what follows, dimensionless lengths are measured in the string scale $\ell_s = 2 \pi \sqrt{\alpha'}$ with Regge slope $\alpha'$.

The K\"ahler form of $X$ may be expanded in the generators $D_i$ of $H^{1,1}(X,\Z)$,
$$
J = t^i D_i \, .
$$
It follows that the volume $\V$ of $X$ may be expressed as \cite{Cicoli:2008va,Baumann:2014nda}
\begin{equation} \label{voldef}
 \V = \frac{1}{6} \int_X J \wedge J \wedge J = \frac{1}{6} \kappa_{ijk} t^i t^j t^k \, .
\end{equation}
Using the Poincar\'e dual four cycles $D_i$ in $H_4(X,\Z)$,\footnote{We denote the Poincar\'e dual pair of forms and cycles using the same notation.} which are divisors of $X$, we define the triple intersection numbers $\kappa_{ijk} \equiv D_i \cdot D_j \cdot D_k$. 
To leading order, the K\"{a}hler potential $K$ is given by
\begin{equation}
K = K_{0} - 2\ln\V \equiv \frac{\hat{K}}{M_{\rm Pl}^2} \, , \label{eq:Ktree}
\end{equation}
where $K_{0}$ encapsulates the $S$ and $U_{a}$ dependence and we have restored the mass dimension in $\hat{K}$.

Further natural quantities are the volumes of the four-cycles $D_i$,
\begin{equation} \label{vol4cycle}
 \tau_i = \frac12 \int_{D_i} J \wedge J = \frac12 \kappa_{ijk} t^j t^k \, .
\end{equation}
These form the complexified K\"ahler moduli, $T_i = \tau_i + i b_i,i=1,\dots,h^{1,1}(X)$, along with the 
shift-symmetric axions 
\begin{equation} \label{axion}
 b_i =  \int_{D_i} C_{(4)} \, ,
\end{equation}
where $C_{(4)}$ is the RR four-form. 

The continuous shift symmetry is perturbatively exact and is only broken by non-perturbative effects. Here it is broken to a discrete symmetry by branes wrapped on the corresponding four-cycles $D_i$. The normalization in (\ref{axion}) gives the periodicity $b_i \to b_i + 1$, such that the generators $D_i$ form a basis of the integral homology.

We consider a number $M$ of stacks of D7- or Euclidian D3-branes. Each stack wraps cycles $D_j$ an integer $Q_{Ij}$ number of times,
\begin{equation} \label{fourcycles}
 D_I = Q_{Ij} D_j\, , \quad Q_{Ij} \in \Z \, , \quad I=1,\dots,M \, .
\end{equation}
Along these branes we can define new K\"ahler moduli generalizing (\ref{vol4cycle}) and \eqref{axion},
\begin{equation}
 \int_{D_I} \left(\frac12 J \wedge J + i C_{(4)} \right)= Q_{Ij} T_j \, .
\end{equation}
The non-perturbative superpotential that obeys the above periodicity is 
\begin{equation}
W = W_0 + \sum_{I=1}^{M} A_I e^{- c_I Q_{Ij} T_j} \equiv \frac{\sqrt{4 \pi}}{g_s^{3/2} M_{\rm Pl}^3}\hat{W} \, , \label{eq:Wnonpert}
\end{equation}
where $W_0$ is the order-one tree-level superpotential, to be discussed in \eqref{WGKP}, $A_I$ are order-one coefficients and
\be \label{NPcoeff}
c_I =
\begin{cases} 2\pi \, , \qquad\; \text{ED3} \, , \\
	2\pi /N_I \, , \quad \text{D7} \, . \\
\end{cases}
\ee
In the former case, a D3-brane wrapped on a four-cycle becomes an instanton and the superpotential describes the instanton effect.
In the latter, a stack of $N_I$ D7-branes gives rise to a $U(N_I)$ gauge theory, and the corresponding superpotential is the well-known one from gaugino condensation, with gauge coupling $\tau_{i}$. 

\begin{figure}[t]
 \begin{center}
\includegraphics[width=5cm]{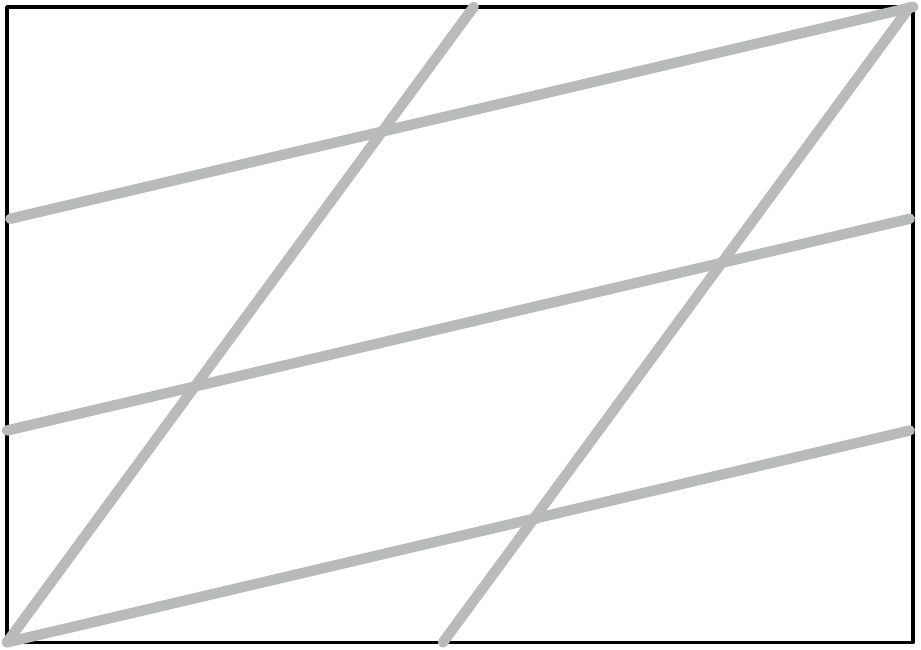}
\end{center}
\caption{Two stacks of branes, wrapping the four cycles $D_I,I=1,2$ differently, are intersecting. The inverse of the length of the fundamental domain segment is proportional to $Q_\xi$ and gives the effective decay constant.}
\end{figure}

Consideration of the D7-gauge anomaly shows that indeed this winding number $Q_{Ij}$ is nothing but the anomaly coefficient (\ref{anomcoeff}). For the D3-worldvolume gauge theory the same interpretation follows, which can be extrapolated to the case of a Euclidian D3-brane. In this brane setup, negative $Q_{ij}$ is also allowed and corresponds to D-branes winding in the opposite direction. To ensure stability of the D-brane setup we need supersymmetry, which in turn requires a special relation between the charges. This relation is also subject to global consistency conditions. In this work, we focus on the dynamics of the axions and leave the complete construction including the Standard Model to future work.

The range of the moduli fields $\tau_i,t^i$ are further restricted because volumes should be positive-definite and the instanton action should be positive. 
For all holomorphic curves, the the K\"aher form should be positive definite,
\begin{equation}
 \int_C J > 0 \,, 
\end{equation}
such that the resulting volumes are non-negative: that is, $J$ should lie in the K\"ahler cone. We may express  this condition in terms of the set of effective curves $C$, by demanding that $C \cdot D \ge 0$ for any divisors $D$ of $X$. This set of curves forms the Mori cone.  Writing the generators of the Mori cone as $C_i$, the K\"ahler cone condition becomes
\begin{equation} \label{Kahlerconecond}
\int_{C_i} J =t^j  \int_{C_i}  D_j \ge 0 \, .
\end{equation}
The conditions on $t^j$ can in turn be translated into conditions on $\tau_i$ through the relation (\ref{vol4cycle}). This not only guarantees the positivity of the volumes $\tau_i$, but restricts their allowed ranges.
 
The leading axion-dependent terms in the scalar potential are
\begin{equation} 
 V \supset - e^{K} K^{j \bar \imath} \sum_{I=1}^{M}\left[  A_I c_I Q_{Ij} e^{-c_I Q_{Ik}T_k}\overline W_0 \partial_{\bar \imath} K + \overline A_Ic_I Q_{I {\bar \imath}}  e^{- c_I Q_{Ik} \overline T_k}W_0 \partial_j K \right] \, .
\end{equation}
Assuming $W_0$ to be $\mathcal{O}(1)$, further terms such as $e^{-c_I Q_{Ij} T_j - c_I Q_{Ij} T^*_j}$ are suppressed, since we are considering the region ${\rm Re} T_i >1$ and $c_I Q_{Ij} = 2\pi Q_{Ij}/N_I \simeq 2\pi \gg 0 $, in which the exponential suppression is sufficiently strong.
Using the relation $K^{j \bar \imath } \partial_{\bar \imath} K =-2 \tau_j, K^{j \bar \imath} \partial_{j} K = -2 \tau_i$, the leading order scalar potential becomes 
\begin{equation} \label{scalarpotential}
 V = 4e^{K} W_0 \sum_{I=1}^{M}  A_I S_I e^{-S_I} \cos(c_I Q_{Ij}b_j) \, ,
\end{equation}
with the instanton actions
\begin{equation} \label{instaction}
 S_I = c_I Q_{Ij} \tau_j \, .
\end{equation}
Note that we may complexify the axion and obtain the above superpotential purely from supersymmetry. 
Thus, it is supersymmetry which is responsible for fixing the instanton action in terms of the K\"ahler moduli. Naturally, the superpotential makes sense if the instanton action is positive-definite. 
Note also that is the leading order expression. Furthermore, to complete the scenario we need moduli stabilization and correlation functions between the various fields, which should be related to supersymmetry breaking and uplifting.

\subsection{Choice of internal manifold} 

The physics of axions is highly dependent on the internal space. Axions are paired with the volume moduli of four-cycles, and thus the geometrical structure of the four-cycles determines the axion potential and decay constants through the superpotential and K\"ahler metric, respectively.   It then stands to reason that an appropriate choice of Calabi--Yau orientifold is necessary for realizing axion alignment.

A general Calabi--Yau orientifold gives rise to multiple $h^{1,1}>1$ axions. In principle, alignment is a restriction on charges as in (\ref{alignment}), and regardless of the sizes of the decay constants, all of these axions can participate in the alignment. Here we will argue that 
\begin{quote}
The axions taking part in the alignment should be those associated with {\em anisotropic bulk four-cycles.} 
\end{quote} 
By bulk cycles we mean those with hierarchically large volumes in the sense of moduli stabilization by means of the Large Volume Scenario (LVS). By anisotropic, we mean that the bulk volume is locally factorizable as a product of those bulk cycles, such that multiple light axions remain after moduli stabilization.
This nontrivial requirement arises because the axions associated with small cycles are decoupled at high energy scales by moduli stabilization. Furthermore, as we will see in section \ref{sec:wgcmultiple}, the weak gravity conjecture favors large cycles with similar sizes. In other words, we consider anisotropic bulk geometry in the near-isotropic limit. The importance of anisotropic geometry has been stressed in many works \cite{Cicoli:2008gp,Cicoli:2016xae,Cicoli:2017axo,Cicoli:2011it}. 

A minimal choice  on which we will focus our analysis in this paper is the well-known K3 fibration over $\mathbb{CP}^1$.
In this case, the bulk volume is determined by two moduli $\tau_{1}$ and $\tau_{2}$,
\begin{equation} \label{fiberdvol}
\V = \alpha\sqrt{\tau_{1}} \tau_2 \, .
\end{equation}
Here the overall normalization $\alpha$ is determined by a concrete realization of the geometry; however, shortly we see that the key physics is independent of $\alpha$. 
One example of this type of geometry can be realized as a degree-12 hypersurface in ${\mathbb C}\P^4_{[1,1,2,2,6]}$ \cite{Cicoli:2008va} under a suitable redefinition.  In our explicit calculations we will mostly refer to this example --- the geometric construction is summarized in the Appendix.

We will stabilize the moduli using Large Volume Scenario. This mechanism relies on  contributions from additional small blow-up cycles. To this end, we also supplement our construction with ``swiss cheese'' geometry by including small internal cycles.
For the case of ${\mathbb C}\P^4_{[1,1,2,2,6]}(12)$, we introduce a further small cycle denoted $\tau_s$. The overall volume is now of the form 
\be \label{volume}
 \V = \alpha (\tau_1^{1/2} \tau_2 - \gamma \tau_s^{3/2}) \, ,
\ee
where $\gamma$ depends on the details of the blow-up.

Other examples can yield similar geometries.  The manifold $\mathbb{CP}^4_{[1,1,2,2,2]}(8)$  obtains a similar form,
\begin{equation}
 \V = \alpha(\tau_1^{1/2} \tau_2 - \gamma_3 \tau_3^{3/2} - \gamma_4 \tau_4^{3/2}) \, ,
\end{equation}
with $\alpha=\sqrt{2}/12$ and $\gamma_3=\gamma_4=2\sqrt2/3$. 
Also the geometry ${\mathbb C}\P^4_{[1,1,1,1,4]}(8)$ with three blow-ups yields a volume \cite{Cicoli:2017axo}
\begin{equation}
 \V = \alpha (\sqrt{\tau_1 \tau_2 \tau_3} - \gamma \tau_s^{3/2}) \, ,
\end{equation}
with $\alpha=\sqrt{2}/4$ and $\gamma=2\sqrt{2}/3$, leading to three large cycles and one small cycle. Although the three manifolds introduced here appear structurally different, all three are K3 fibrations over a two-base and lead to the same general structure,
\begin{equation}
 \V = n t^a \tau_a - \alpha \sum_{\tau_i \text{ small}} \gamma_{i} \tau_{i}^{3/2} \quad \text{($a$ not summed)} ,
\end{equation}
where $\tau_a$ is the volume of the K3 fiber, $t^a$ is that of the two-base and $n$ is the integer power to which $t^a$ appears in the bulk volume, $\V \propto (t^a)^n$.  There are also requirements on the small cycles.  In particular, in order to have a stable minimum, at least one of them should be a blown-up cycle resolving a singularity \cite{Cicoli:2008va}.

One could also consider manifolds with more cycles, but in the present work we limit our discussion to small $h^{1,1} \lesssim 4$. This is because the K\"ahler cone condition restricts the range of moduli space, prohibiting large decay constants \cite{Denef:2004dm}; moreover, with many cycles the cone becomes very narrow, yielding large cycle volumes $\tau_i \sim (h^{1,1})^p$ \cite{Demirtas:2018akl}. Although the KNP mechanism works well regardless of the axion decay constants, for consistent moduli stabilization other cycles should be sufficiently small, such that we may expect their relatively heavy axions to decouple. 

Note that in all of the above geometries, the volumes $\tau_i$ are not those of the cycles of integral cohomology $H^{1,1}(X,\Z)$. We need a suitable redefinition of the diagonal form of the volume, which depends on the details of the construction. This imposes an important constraint on the range of the K\"ahler parameter $t^i$ to satisfy the K\"ahler cone condition (\ref{Kahlerconecond}).  For example,
\begin{align}
 {\mathbb C}\P^4_{[1,1,2,2,6]}(12)&: \tau_2 \ge \frac{4}{3} \tau_1 >0,\label{ourkahlercone} \\
 {\mathbb C}\P^4_{[1,1,2,2,2]} (8)&: \tau_2 \ge 4 \tau_1 >0,\ \tau_3 \ge \tau_4 \ge \frac{1}{2} \tau_1 >0, \\ 
 {\mathbb C}\P^4_{[1,1,1,2,4]} (8)&: \frac{2 \tau_s}{\tau_i} < \frac{\tau_j}{\tau_k} < \frac{\tau_i}{2 \tau_s}, \quad  i\ne j\ne k. 
\end{align}

Finally, the Large Volume Scenario makes use of an ${\cal O}(\alpha^{\prime 3})$ correction in order to realize the large-volume minimum, which should induce a positive contribution to the potential.  We will see that requiring this term to be positive implies that the Euler number should be negative, $ \chi(X) \equiv 2h^{1,1} - 2h^{2,1} < 0.$ 

\subsection{Axion decay constants}

The axion decay constants are determined by the geometry and encoded in the K\"aher metric.
The K\"ahler metric $K_{i \bar \jmath} \equiv \partial^2 K / (\partial T_i \partial T_j^*) = 4 \partial^2 K / (\partial \tau_i \partial \tau_j) $ is a real, symmetric matrix, giving the kinetic terms
\begin{equation} \label{kinetic}
{\cal L}_{\rm kin} = \sum_{i,j} K_{i \bar \jmath} \partial_\mu T_i \partial^\mu T_j^* \, .
\end{equation}

If the K\"ahler metric is diagonal, each entry is the square of the axion decay constant, in the normalization that the scalar fields have periodicities $2\pi$,
$$ 
{\cal L}_{\rm kin} = \sum_i\left[\frac{f_i^2}{2} \partial_\mu (2\pi b_i) \partial^\mu (2\pi b_i) + \frac{f_i^2}{2} \partial_\mu (2\pi \tau_i) \partial^\mu (2\pi \tau_i)\right] \,.
$$
(Note the different normalization from $\vartheta$ in Section \ref{sec:aligned}).
Comparing these two expressions, we may read off the decay constants,
$$
f_i^2 =  \frac{1}{2\pi^2} \langle K^{\rm diag}_{i\bar \imath} \rangle \quad (\text{no summation)} \, .
$$
For a non-diagonal K\"ahler metric, we may define the matrix-valued decay constant ${\bf f}$ as in (\ref{decayconstmatrixnorm}),
\begin{equation} \label{decayconstmatrix}
  {\bf f}_{ki} {\bf f}_{kj} = \frac{1}{2\pi^2} \langle K_{i\bar \jmath} \rangle \, . 
\end{equation} 

The moduli dependence of decay constants can be understood by considering a generic volume of the form\footnote{We assume three or fewer bulk four-cycles. Calabi--Yau manifiolds in general are non-simplicial and we cannot express the volume simply in terms of $\tau_i$s.}
\begin{equation} \label{volansatz}
 \V = \alpha (\tau_1^{p_1} \tau_2^{p_2} \tau_3^{p_3} + \beta \tau_1^{q_1} \tau_2^{q_2} \tau_3^{q_3}) \equiv \alpha(\V_f + \beta \V_s) \,.
\end{equation}
Here due to the intersection structure (\ref{voldef}), the powers $p_i,q_i$ can take only non-negative half-integer values. Noting that $\tau_i$ are the volumes of four-cycles, the powers must additionally satisfy $p_1+p_2+p_3=q_1+q_2+q_3 = 3/2 = \dim_{\mathbb C} X/2$, with $dim_{\mathbb C} X = 3$ the complex dimension of $X$.
In the simple case $p_1=p_2=p_3=1/2, \beta=0$, \eqref{volansatz} describes the volume of a complex threefold as a product of three Riemann surfaces. 

First let us consider generic $p_1,p_2,p_3$, with $\beta=0$. 
In most cases, this structure requires diagonalization.
The K\"ahler potential becomes 
$$ K = K_0 - 2 \log \alpha -2 p_1 \log \tau_1 -2 p_2 \log \tau_2 -2 p_3 \log \tau_3 \, ,
$$
yielding the diagonal K\"ahler metric
$$
 K_{i \bar \jmath} =\frac{1}{2} {\rm diag} \left( \frac{p_1}{ \tau_1^2}, \frac{p_2}{\tau_2^2},\frac{p_3}{\tau_3^2} \right)\,.
$$
Thus the decay constants take the form
\begin{equation} \label{axiondecayconsts}
 f_i = \frac{\sqrt{ p_i} }{2 \pi  \tau_i}M_{\rm Pl} \, ,
\end{equation}
depending on the powers of $\tau_i$ appearing in the volume, but not the overall coefficient, due to the logarithmic dependence of the K\"ahler metric. Although the values of the moduli $\tau_i$ correspond to cycle volumes in units of the string length, the K\"ahler potential is measured in terms of the Planck scale \cite{Conlon:2005ki}. The validity of the supergravity description requires $\tau_i >1$, so as expected, string theory naturally yields dimensionful parameters below the Planck scale.

The correction to the volume as in the second term of (\ref{volansatz}) introduces off-diagonal components in the K\"ahler metric,
\begin{align}
K_{i\bar \imath} &= \frac{\alpha^2}{2 \tau_i^2 \V^2} \left(p_i \V_f^2 - \left((p_i-q_i)^2-p_i-q_i\right) \beta \V_f \V_s + q_i \beta^2 \V_s^2 \right) \,, \\
K_{i\bar \jmath} & =  -\frac{\alpha^2 \beta}{2 \tau_i \tau_j \V^2} (p_i-q_i)(p_j-q_j) \V_f \V_s \, , \quad i \ne j\,.
\end{align}
In the case where $p_i \ne q_i$ and $p_j \ne q_j$, this implies that the geometry should have a fibered structure along the $i,j$ directions.

If we interpret $\V_s$ as a small correction (although it is not always the case) then we may approximate $\V_f \gg \beta \V_s$,  which simplifies the K\"ahler metric,
\begin{align}
K_{i\bar \imath} &= \frac{1}{2 \tau_i^2 } \left(p_i - \left((p_i-q_i)^2-p_i-q_i\right) \frac{\beta  \V_s }{\V_f}+ q_i \left(\frac{\beta \V_s}{ \V_f}\right)^2 \right) \, , \label{subleading1} \\
K_{i\bar \jmath} & = - \frac{1}{2 \tau_i \tau_j } (p_i-q_i)(p_j-q_j) \frac{\beta \V_s}{\V_f} \, .\label{subleading2}
\end{align}
Note that the off-diagonal components $K_{i\bar \jmath}$ are subleading in $\beta \V_s/\V_f$, and furthermore vanish if the volume is completely factorized in $\tau_i$ or $\tau_j$. Moreover, in the $\beta\V_s/\V_f \to 0$ limit the K\"ahler metric becomes diagonal and reproduces the previous result (\ref{axiondecayconsts}). We may generalize this argument to cases with additional terms in the expression for the volume.

\begin{figure}[t!]
 \begin{center}
\includegraphics[width=7cm]{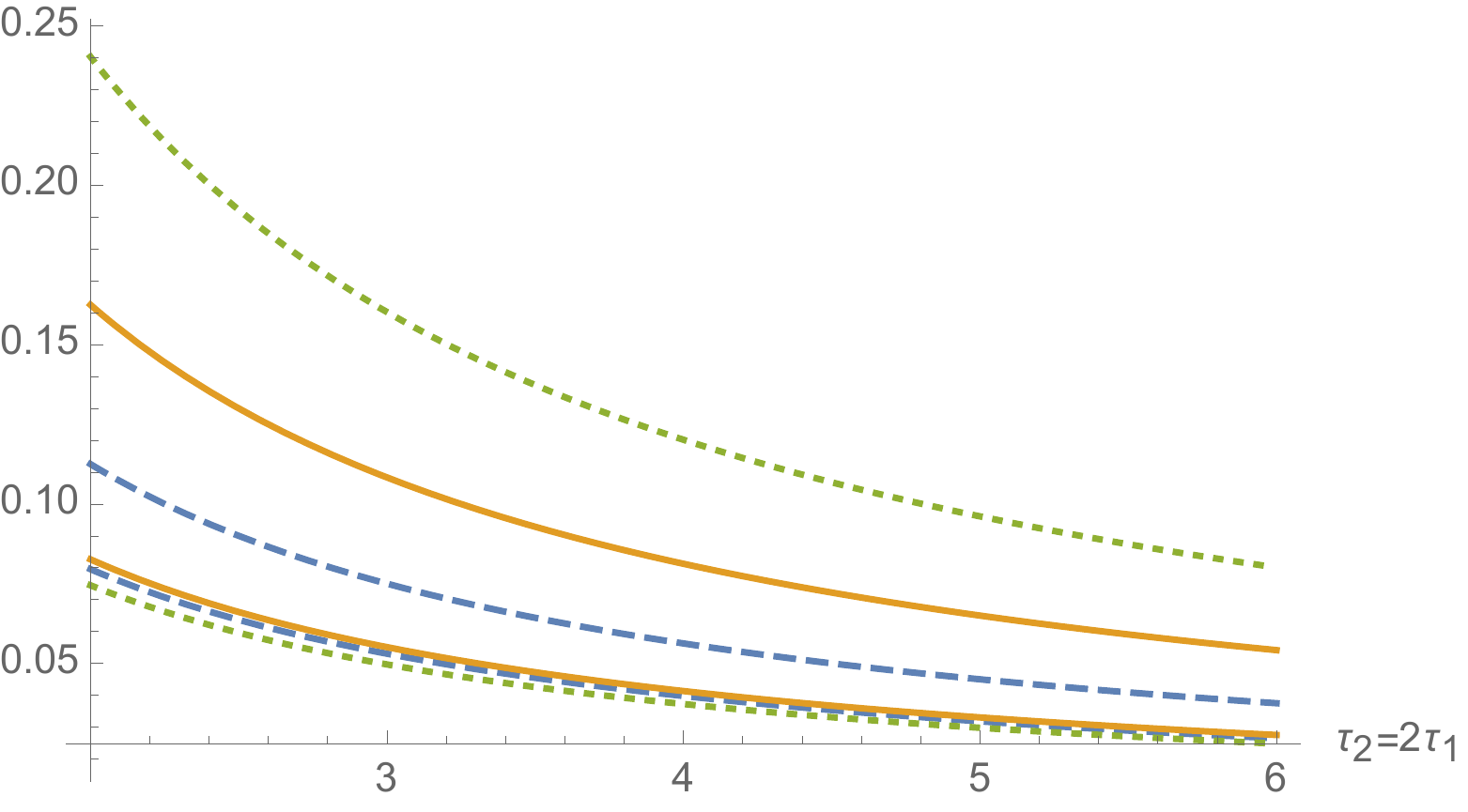}
\includegraphics[width=0.45\textwidth]{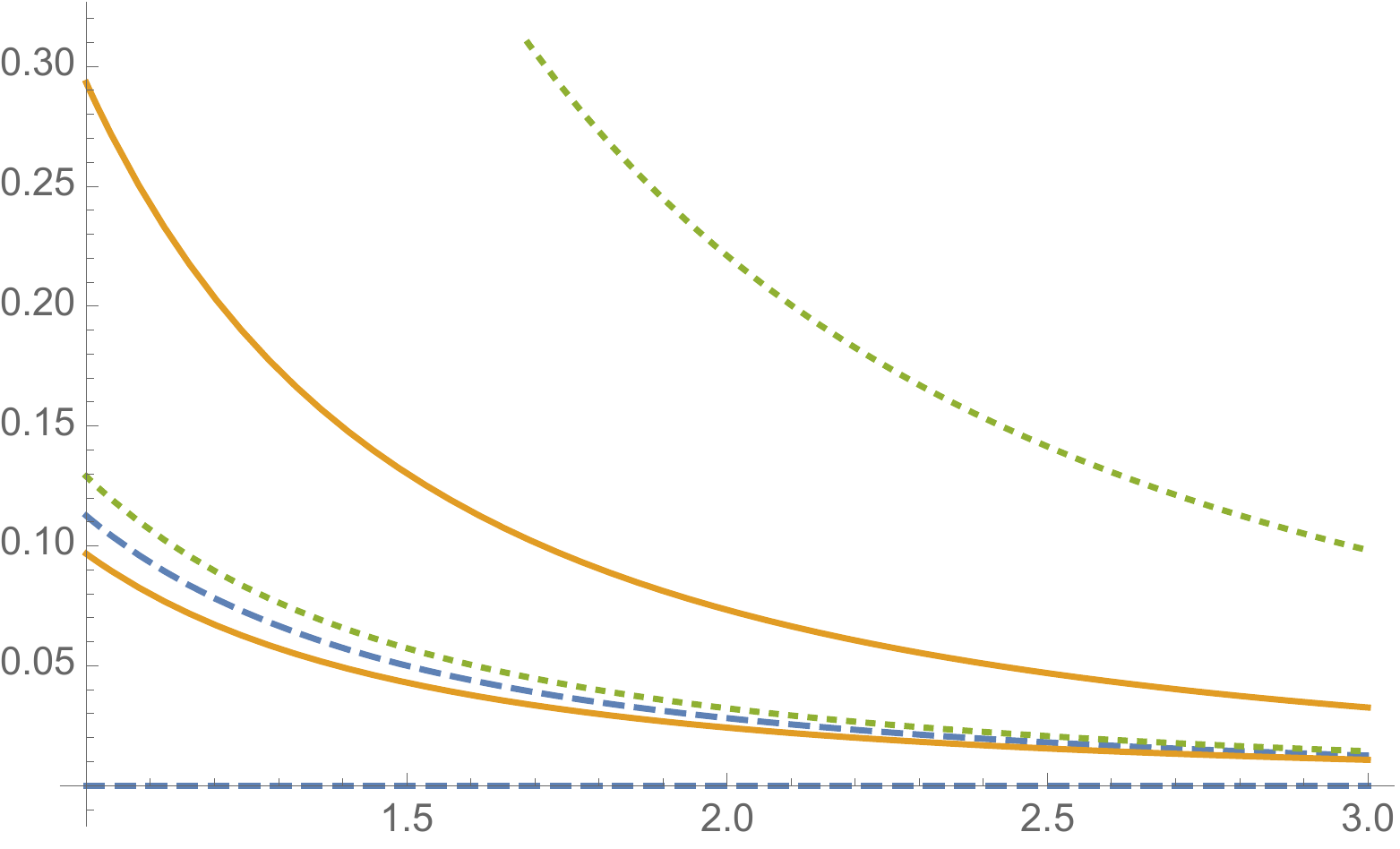}
\end{center}
\caption{The ``decay constants'', defined as the square-roots of the eigenvalues of the K\"ahler metric divided by $2\sqrt 2  \pi,$ for two light axions, assuming a volume of the form (\ref{volansatz}). \textit{Left:} Two-axion scenario, cases $\beta=0$ (solid)$,-2/3$ (dashed)$,-1$ (dotted), with $(p_1,p_2,p_3)=(1/2,1,0)$ and $(q_1,q_2,q_3)=(3/2,0,0)$. We fixed the K\"ahler moduli to satisfy the relation $\tau_2=2\tau_1$; general behaviors are similar for other relations. \textit{Right:} Three-axion scenario, cases for $(p_1,p_2,p_3)=(1/2,1,0)$ and $(q_1,q_2,q_3)=(0,0,3/2)$. Solid, dashed and dotted curves correspond to $(\beta,\tau_1/\tau_2,\tau_1/\tau_3) = (0,1/2,2), (-1,1/2,2),(-1,1/2,1)$, respectively. In this case $\tau_3$ may be interpreted as a small cycle, which decouples due to its heavier mass as a result of having a smaller instanton action.
} \label{fig:decayconsts}
\end{figure}

We draw the ``decay constants'', defined as the eigenvalues of the K\"ahler metric divided by $2\sqrt 2  \pi,$ for the case of two light axions in Fig. \ref{fig:decayconsts}. Different curves correspond to $\beta=0,-2/3,-1$,  respectively. We chose the volume to be of the form (\ref{volansatz}) with $(p_1,p_2,p_3)=(1/2,1,0),(q_1,q_2,q_3)=(3/2,0,0)$ and K\"ahler moduli satisfying the relation $\tau_2=2\tau_1$. General behaviors are similar for other relations, that is, decreasing as $\tau_i$ with a negative power.
The dominant behavior (\ref{axiondecayconsts}) is largely unmodified by subleading contributions (\ref{subleading1})--(\ref{subleading2}). 

While the decay constants are determined by the K\"ahler potential, their effects in the axion potential (\ref{scalarpotential}) always appear in the combination
\begin{equation}
   \frac{2\pi Q_{Ij}  b_j}{N_I f_j} \, ,
\end{equation}
 with the charge and, in the case of gaugino condensataion, the ranks $N_I$ of the condensate groups. This effectively redefines the charges $Q_{Ij} \to Q_{Ij}/N_I$.
 Therefore, in the case of two-axion alignment, the effective decay constant becomes
 \begin{equation}  \label{feffN}
 f_{\rm eff} =  N_2 \frac{\sqrt{f_1^2 Q_{12}^2 + f_2^2 Q_{11}^2}}{|Q_{21} Q_{12} - Q_{11} Q_{22}|} \, .
 \end{equation}
Again, we should ensure that there is no tunneling between degenerate branches of vacua. We may conclude that modification of the charges does not give rise to any  naturalness problem in the axion decay constant.

\subsection{The Weak Gravity Conjecture for an axion with a dilaton}

The alignment mechanism appears to allow for a trans-Planckian axion decay constant in a certain basis of field space. The alignment restores the shift symmetry, which is a global symmetry, but global symmetries are believed to be violated by quantum gravity effects. For example, if we consider a black hole, a globally charged object cannot preserve its charge because Hawking radiation cannot carry global charge. This implies that quantum gravity processes involving a black hole violate the global symmetry. A more general principle is quantitatively formulated as the weak gravity conjecture (WGC) \cite{ArkaniHamed:2006dz}. 

For the case of a $U(1)$ gauge theory, 
\begin{equation} 
{\cal L} =  \frac{1}{2} M_{\rm Pl}^2 R - \frac{1}{4g^2}  F_{\mu\nu} F^{\mu\nu} \, , 
\end{equation} 
the weak gravity conjecture is the requirement that there exists a charged particle with mass $m$ and gauge charge ${\cal Q}=qg$, where 
$q$ is the quantized charge, such that 
\begin{equation}  
 m  \leq qg\sqrt{2} M_{\rm Pl} \, . 
\end{equation} 
This originates from the condition for extremal black hole decay, imposing the charge-to-mass ratio bound
\begin{equation} 
 \frac{qgM_{\rm Pl}}{m} \geq \left(\frac{{\cal Q} M_{\rm Pl}}{M}\right)_{\rm extremal\, BH} = \frac{1}{\sqrt{2}} \, .
\end{equation} 
The extremal condition is consistent with the force-free condition for extremal objects.
This criteria can be modified in the presence of a dilatonic coupling for the gauge field, as 
\begin{equation} 
{\cal L} =  \frac{1}{2} M_{\rm Pl}^2 R +\frac{1}{2} (\partial_\mu \phi)(\partial^\mu \phi)  - \frac{e^{-\alpha\phi}}{4 g_0^2} F_{\mu\nu} F^{\mu\nu} \, .
\end{equation} 
Here the gauge coupling depends on the field $g = g_0 e^{\alpha \phi/2}$, so we expect a nontrivial black hole solution. As given by \cite{Garfinkle:1990qj}, an extremal dilatonic black hole gives the following modified equality: 
\begin{equation} 
\left(\frac{{\cal Q} M_{\rm Pl}}{M}\right)_{\rm extremal\, BH}= \sqrt{\frac{1}{2} + \frac{\alpha^2}{2}} \, .
\end{equation} 
This is also the balance condition for extremal objects between the repulsive $U(1)$ gauge force and the attractive force by gravity and the dilaton. The corresponding WGC can be inferred as \cite{Heidenreich:2015nta,Palti:2017elp} 
\begin{equation} \label{WGCphoton}
\frac{qg M_{\rm Pl}}{m} \geq \sqrt{\frac{1}{2} +\frac{\alpha^2}{2}} \, .
\end{equation}

Including the dilatonic coupling, this relation can easily be generalized to higher dimensional objects in general spacetime dimensions, as long as we can identify the relevant extremal macroscopic objects. 
However, it is not clear what the corresponding extremal objects are in the case of axions.
By naive analogy with the $U(1)$ case, we may replace $(m,q)$ with $(S,Q)$. Here $Q$ is the anomaly coefficient in (\ref{anomcoeff}), interpreted as a quantized instanton charge, measured in units of the axion decay constant $1/f$, and $S$ is the corresponding instanton action. Then the WGC for axions can be written as 
\begin{equation} \label{oneaxionWGC}
\frac{ Q M_{\rm Pl}}{f S }  \geq \left( \frac{ {\cal Q} M_{\rm Pl}}{S}\right)_{\text{extremal}} \, ,  
\end{equation} 
where ${\cal Q}$ is the charge and $S$ is the instanton action of an extremal object. 
Naive comparison of (\ref{WGCphoton}) to the axion case implies that the right-hand side of (\ref{oneaxionWGC}) becomes $\alpha/2$. 

On the other hand, 
if we consider axions in string theory, the dilatonic dependence in the axion coupling to the instanton arises naturally.
In a specific axion model for $T=\tau + ib$ with $b\simeq b+1$, and $\hat{K}/M_{\rm Pl}^2 = - 2  \ln \tau^p$, we find
\begin{equation}\label{dilaton-axion}
\begin{split}
{\cal L} &=\frac{1}{2} M_{\rm Pl}^2 \left(R + p\frac{(\partial_\mu\tau)^2}{\tau^2} +p\frac{(\partial_\mu b)^2}{\tau^2} \right) \\
&= \frac{1}{2} M_{\rm Pl}^2 R+ \frac{1}{2}(\partial_\mu\phi)^2 + \frac{1}{2} \left(\sqrt{p}M_{\rm Pl}e^{-\phi/(\sqrt{p}M_{\rm Pl})}\right)^2 (\partial_\mu b)^2 \\
&= \frac{1}{2} M_{\rm Pl}^2 R +  \frac{1}{2} (\partial_\mu \tilde{\phi})^2 + \frac{1}{2} (2\pi f_0)^2 e^{-\alpha\tilde{\phi}/M_{\rm Pl}} (\partial_\mu b)^2 \, .
\end{split}
\end{equation}
The second line is obtained by redefining $\phi = \sqrt{p}M_{\rm Pl}\ln(2\pi\tau)$, and in the third line we have expanded $\phi = \phi_0 + \tilde{\phi}$ around its vacuum expectation value, $\phi_0 \equiv \sqrt{p}M_{\rm Pl}\ln\tau_{0}$, with $f_0 = \sqrt{p}M_{\rm Pl}/(2\pi\tau_0)$.  This fixes the dilaton charge in the third line of (\ref{dilaton-axion}) as
\begin{equation}
 \alpha=2\sqrt{\frac{1}{p}} \geq 0 \, .
\end{equation}
Thus, the field-dependent axion decay constant is given by (\ref{axiondecayconsts}). 
The  corresponding instanton action with quantized charge $Q$ in units of $1/f = e^{\alpha\tilde{\phi}/(2M_{\rm Pl})}/f_0$  becomes that of (\ref{instaction}). 
From these we obtain the relation 
\begin{equation}\label{BPS}
\left(\frac{ Q M_{\rm Pl}}{f S}\right)_{\rm string\, axion} = \sqrt{\frac{1}{p}} =\frac{\alpha}{2} \,.
\end{equation} 
This is a consequence of supersymmetry and is also related to the force balancing condition for the corresponding dual object, strings, in the limit of vanishing scalar potential. 
Note that even if the saxion is stablized by its potential and thus becomes massive by supersymmetry-breaking effects, the relation (\ref{BPS}) still holds as a remnant of the underlying supersymmetric structure.

If the saxion is totally decoupled, and we consider only the axion and gravitational interaction, 
the Giddings--Strominger (GS) solution for the euclidean wormhole (WH)  \cite{Giddings:1987cg} can provide an interesting implication. For a wormhole solution with instanton action $S$ and PQ charge ${\cal Q}=Q/f$ carried by the wormhole throat, the following relations hold \cite{Hebecker:2016dsw,Andriolo:2020lul}:
\begin{equation} 
\left( \frac{ {\cal Q} M_{\rm Pl}}{S}\right)_{\rm WH(GS)}= \frac{4}{\sqrt{6}\pi} \, .
\end{equation} 
One can also allow coupling to the dilaton as in the Lagrangian of the third line in (\ref{dilaton-axion}).
In such a case, assuming that the dilaton mass is vanishingly small around the wormhole throat, the euclidean wormhole solution only exists for $\alpha < 2\sqrt{2/3}$. In this regime \cite{Hebecker:2016dsw,Andriolo:2020lul}, 
\begin{equation} 
\left( \frac{ {\cal Q} M_{\rm Pl}}{S}\right)_{\rm WH} = \frac{\alpha}{2\sin\Big(\frac{\pi\alpha}{4\sqrt{2/3}}\Big)} \, .
\end{equation} 
One can take the limit of $\alpha\to 2\sqrt{2/3}$ to obtain the maximum value of the charge-to-mass ratio. Then
\begin{equation} \label{axionWGC} 
\left(\frac{ {\cal Q} M_{\rm Pl}}{S}\right)_{{\rm extremal\, WH}} =\left.\frac{\alpha}{2}\right|_{\alpha\to 2\sqrt{\frac{2}{3}}} = \sqrt{\frac{2}{3}} \, ,
\end{equation}
which is also dubbed an {\it extremal gravitational instanton} \cite{Bergshoeff:2004fq,Heidenreich:2015nta}, since the wormhole throat size is vanishing and becomes singular, so this object can really absorb the PQ charge without need of a baby universe. 

Interestingly, the RHS of (\ref{axionWGC}) is coincident with (\ref{BPS}) in the no-scale limit $2p=3$. This is naturally the case if we interpret $2p$ as the complex dimension of the internal manifold. We can also show that this result holds for multiple axions if they are associated with the K\"ahler moduli of four-cycles (see examples below).
For the extremal objects entering the axion WGC, one possible choice is to consider the string axions.

\subsection{The WGC for multiple axions} \label{sec:wgcmultiple}

Generalization to the case of multiple axions can be performed along the lines of the similar generalization of the particle WGC to the case of multiple $U(1)$ gauge bosons. 
In the string axion case, 
owing to the off-diagonal elements of $K_{i \bar \jmath}$, we have a generalized matrix-valued decay constant ${\bf f}$ as in (\ref{decayconstmatrix}). For the instanton with its $S_I$  and the quantized charges $Q_{Ik}$,  the instanton charge vector is defined as 
\begin{equation}
{\cal Q}_{Ii} = {\bf f}^{-1}_{ik} Q_{Ik} \, .
\end{equation}  
Thus, the covariant generalization of the above WGC is straightforward.

First of all, we can show that if a general decay constant ${\bf f}$ is obtained from the K\"ahler potential (\ref{eq:Ktree}) and 
the internal volume (\ref{voldef}), the following inequality is satisfied for arbitrary charges $Q_{Ii}$:
\begin{equation} \label{theinequality}
 \frac{\sqrt{{\cal Q}_{Ii} {\cal Q}_{Ii}}  M_{\rm Pl}}{|2\pi  Q_{I i}\tau^i | }\geq \sqrt{\frac{2}{3}} \quad 
\textrm{(no sum over $I$)} \, .
\end{equation} 
Note that $\sqrt{2/3}$ in the right-hand side of (\ref{theinequality}) is the coefficient of the inequality in the single axion case with $p=3/2$ (\ref{BPS}). For multiple axions,
this inequality is saturated if the moduli field values are aligned with respect to the charges as
\begin{equation} 
\frac{\partial K}{\partial \tau^i } \propto Q_{I i} \, .
\end{equation} 
Recall that the instanton actions are given in (\ref{instaction}) as
\begin{equation} 
S_I =c_I Q_{Ii}\tau^i \, , 
\end{equation} 
which are required to be positive.
Considering the charge-to-mass ratio for the axion, 
\begin{equation} \label{zvector}
 \mathbf{z}_I =  \frac{ {\cal Q}_{Ii} M_{\rm Pl}}{S_I } \hat e_i \, , 
\end{equation}
the inequality can be written as 
\begin{equation} \label{ctomass}
||\mathbf{z}_I||_{\rm string\, axion} \ge  \sqrt{\frac{2}{3}} \equiv \r \, ,
\end{equation} 
for any charge-to-mass ratio vector.  Hereafter we call the saturated value $\r$.

Furthermore, one can also consider the Euclidean wormhole solution for multiple-axion models.
For Giddings--Strominger wormhole solutions, the norm of the charge-to-mass ratio vector is fixed as 
\begin{equation} 
||\mathbf{z}_I||_{\rm WH(GS)} = \frac{4}{\sqrt{6}\pi}
\end{equation} for any direction of the charge-to-mass ratio vector \cite{Montero:2015ofa}, so the allowed domain is the sphere in the charge-to-mass ratio vector space. 
We are also interested in the extremal wormhole solution for multiple axion-dilaton systems. 
In the case of the single axion-dilaton model, the extremal wormhole solution can be obtained in the no-scale limit ($p=3/2$).
A natural conjecture is then that the multiple-axion-dilaton kinetic metric in the no-scale limit (i.e. where ${\bf f}$ is obtained from (\ref{eq:Ktree}))  without potentials also gives the extremal wormhole solution with vanishing wormhole throat size.  This turns out to be true \cite{ArkaniHamed:2007js}, and the Euclidean action for given quantized charges $Q_{Ii}$ becomes  
\begin{equation}
S_I = 2\pi Q_{Ii}\tau^i \quad \Rightarrow \quad ||\mathbf{z}_I||_{\rm extremal\, WH}  
= \frac{\sqrt{{\cal Q}_{Ii } {\cal Q}_{I i}}M_{\rm Pl}}{|2\pi Q_{I i}\tau^i|} \, ,
\end{equation}
the same as in the string axion case.  The domains of the charge-to-mass ratio vectors for string instantons and extremal wormholes 
is a hyperplane for given saxion values rather than a sphere, since $|\mathbf{z}_I\cdot (2\pi {\bf f}_{ij} \tau^j)| = 1$. 
However, it is also of note that as we approach the center of the extremal wormhole ($r\to 0$), the solutions of $\tau^i(r)$ tend towards the values that saturate the inequality (\ref{ctomass}), i.e.  $\partial K/\partial \tau^i \propto Q_{I i}$ for $r\to 0$.

The WGC for multiple axions can be expressed as the existence of a {\it convex hull} constructed from the charge-to-mass ratio vectors, which contains the domain of the allowed $(\mathbf{z}_I)_{\rm extremal}$  \cite{Cheung:2014vva,Brown:2015iha}. 
The explicit value of $||\mathbf{z}_I||_{\rm extremal}$ for each direction of $\mathbf{z}_I$ may not be clearly specified because the form of the domain is easily deformed by the existence of the dilatontic partners. 
If we consider Giddings--Strominger wormholes, the charge-to-mass ratio vectors span a sphere with a radius $4/\sqrt{6}\pi$. 
If we consider string axions as the objects giving rise to $(\mathbf{z}_I)_{\rm extremal}$, the extremal bound becomes a hyperplane for given saxion values, whose minimal norm is $\r$ as (\ref{ctomass}).

Interestingly, the K\"ahler moduli enter into both the axion decay constants and the instanton actions. 
In the context of type II string compactifications, the charge-to-mass ratio vector \eqref{zvector} 
is {\em invariant under rescalings} of either the charges or the K\"ahler moduli (see for example \eqref{thirdWGCvec} in section \ref{sec:restrictedparams}).  Furthermore, the contributions from group-theoretical factors $N_I$ also cancel.
This implies that the CHC constrains the ratios $\tau_i/\tau_j$ of the four-cycles in the compact geometry.

In order to more concretely understand the phenomenological implications of the WGC as it applies to the inflationary dynamics of axions with frozen saxions, 
we consider a rather simplified convex hull condition (CHC): (i) the extremal domain of the charge-to-mass ratio vector is given by the sphere with the radius $\r$, (ii) the instantons used to construct the convex hull contribute to the scalar potential of the axions via SUSY breaking effects. 
In fact, the relationship between the instantons satisfying the convex hull condition and the scalar potential is ambiguous because of fermionic zero modes. Nevertheless, one can assess the possible implications from the distribution of allowed instanton actions which may contribute to the scalar potential.  

While each charge-to-mass ratio vector of an instanton may have a norm greater than the value of $||\mathbf{z}_I||_{\rm extremal}$ along its direction, the convex hull spanned by the aligned-instanton charge-to-mass ratio vectors may not necessarily cover the allowed extremal domain.
In the alignment limit, the two vectors are almost parallel so the resulting convex hull becomes line-like and cannot embrace the circle.
A suggested remedy \cite{Rudelius:2015xta} is that we may introduce another instanton giving rise to a third axion potential as follows. Letting its axion charges be $Q_{31},Q_{32}$, the resulting vector $\mathbf{z}_3$ is required to be almost orthogonal to the original two vectors $\mathbf{z}_1$ and $\mathbf{z}_2$, 
\begin{equation} \label{CHCorthogonality}
\frac{\mathbf{z}_I \cdot \mathbf{z}_3}{|\mathbf{z}_I||\mathbf{z}_3|} 
=  \frac{Q_{31} Q_{I1} f_2^2 +Q_{32} Q_{I2} f_1^2} {\prod_{I'=I,3} \sqrt{(f_2 Q_{I'1})^2 +(f_1 Q_{I'2})^2}}  \ll 1, \quad I=1,2,
\end{equation}
such that the resulting convex hull contains the unit circle. These parameters come from the corresponding instanton potential,
\begin{equation}
\Delta V =-  \lambda_3^4 e^{-S_3} \cos \left(\frac{Q_{31}}{f_1} \phi_1 +\frac{ Q_{32}}{f_2} \phi_2\right) \, ,
\end{equation}
and the resulting vector ${\bf z}_3$ generates a new vertex for the convex hull such that it embraces the circle.
In the orthogonal case (\ref{CHCorthogonality}), the potential written in the KNP basis reduces to
\begin{equation} \label{thirdpotential}
\begin{split}
 \Delta V &=  -
 \lambda_3^4 e^{-S_3} \cos \left( \frac{Q_{3\xi}}{f_\xi} \phi_\xi  + \frac{Q_{3\psi}}{f_\psi} \phi_\psi \right) \,, \end{split}
\end{equation}
where 
\begin{equation}
Q_{3\xi} = \frac{Q_{31} Q_{12} - Q_{11} Q_{32}}{\sqrt{Q_{11}^2+ Q_{12}^2}}\, ,\qquad
Q_{3\psi} = \left(\frac{f_1^2 Q_{12} Q_{32} + f_2^2 Q_{11} Q_{31}}{f_1^2 Q_{12}^2 + f_2^2 Q_{11}^2} \right) Q_{1\psi} \, .
\end{equation}
In order not to ruin the small curvature of the original axion potential, we need a large enough instanton action, $S_3 \gg 1$. Note that although the corresponding potential has a large coefficient $S_3$, it is drowned out by the suppression by the exponential of $-S_3$.  Later we will see that this condition $S_3 \gg 1$ is achieved naturally.\footnote{Owing to the relatively small effective decay constant of the potential (\ref{thirdpotential}), it can generate a modulation of the inflaton potential depending on the size of $S_3$. Such a possibility provides a wider range of inflation parameters compared to those of vanilla natural inflation  \cite{Kappl:2015esy,Choi:2015aem}.}

\section{Testing aligned natural inflation} \label{sec:explicitmodel}

In this section, we take concrete values of the moduli fields and compare them with the real observation from cosmology.  After briefly reviewing the Large Volume Scenario, we discuss how it leads to a generic prediction of light axions, which obtain masses only through bulk-volume-suppressed non-perturbative effects \cite{Balasubramanian:2005zx}, we present a low-energy effective potential for the light axion, highlighting the parametric dependence of the axion mass on the underlying geometry.  The decay constants are constrained by string geometry, supersymmetry and the weak gravity conjecture.

\subsection{Moduli stabilization I: The overall volume and the small cycles}

\begin{table}[t]
\begin{center}
\begin{tabular}{cccc}
\hline \hline
Stack & Slices & Supporting cycle & Condition   \\
\hline
1 & $N_1$ & $ Q_{11} D_1+Q_{12} D_2 $ & $Q_{11} Q_{12} < 0$ \\
2 & $N_2$ & $ Q_{21} D_1 +Q_{22} D_2$ & alignment (\ref{alignment}) \\
3 & $N_3$ & $ Q_{31} D_1+ Q_{32} D_2$ & $Q_{31}\ge 0, Q_{32}\ge 0$, orthogonal (\ref{CHCorthogonality})  \\
s & $N_s$ & $D_s$ & \\
\hline
\end{tabular}
\caption{Axionic brane configuration. We choose D7-branes such that we can have multiple slices, with $N_I$ branes in each stack. } \label{t:braneconfig}
\end{center}
\end{table}

Now we turn to moduli stabilization.
The first step is to turn on background fluxes \cite{Giddings:2001yu}, which induces a superpotential \cite{Gukov:1999ya}
\begin{equation}
W = \frac{1}{\ell_s^2} \int_{X} G_{3}\wedge\Omega \, , \label{WGKP}
\end{equation}
where $\Omega$ is the invariant $(3,0)$-form of $X$ and $G_{3} = F_{3} - SH_{3}$ is the $SL(2,\mathbb{C})$-covariant three-form field strength.  This induces an F-term scalar potential which is minimized at $\rD_{S}W = 0$ and $\rD_{a}W = 0$, thus fixing the axio-dilaton and complex structure moduli supersymmetrically.  The remaining F-term scalar potential for the K\"{a}hler moduli takes the form
\begin{equation}
V_{\rm F} = e^{K/M_{\rm Pl}^2}\left[K^{i\bar{\jmath}}\rD_{i}W\rD_{\bar{\jmath}}\overline{W} - \frac{3}{M_{\rm Pl}^2}|W|^{2}\right] \, , \label{eq:scalarKahler}
\end{equation}
where $\rD_{i}W \equiv \partial_{i}W + K_{i}W$, $K_{i} \equiv \p_{i}K$ and $K^{i\bar{\jmath}} = K_{\bar{\imath}j}^{-1}$.

At tree level the effective action has no-scale structure, i.e. $V_{\rm F}$ in  \eqref{eq:scalarKahler} vanishes independent of $T_{i}$.  Thus, in order to stabilize the K\"{a}hler moduli we must include subleading corrections.  There are two types of contributions which combine to realize the Large Volume Scenario.  First of all, the K\"{a}hler potential may receive perturbative corrections.  The leading contribution arises at $\O(\alpha'^{3})$ in the 10-dimensional supergravity action and modifies the K\"{a}hler potential to the form
\begin{equation} 
	K = K_{0} - 2\ln\left(\V + \frac{\xi}{2g_{s}^{3/2}}\right) \, , \qquad \xi \equiv -\frac{\chi(X)\zeta(3)}{2(2\pi)^{3}} \, , \label{eq:Kpert}
\end{equation}
where $\chi(X)$ is the Euler number of $X$. Hereafter we will use $\hat \xi \equiv \xi g_s^{-3/2} $.

Secondly, the superpotential can receive non-perturbative corrections from brane instantons.
We assume that we are working with a volume of the form \eqref{volume},
$$
\V = \alpha (\tau_1^{1/2} \tau_2 - \gamma \tau_s^{3/2}) \, ,
$$
in the regime where $\tau_1, \tau_2 \gg \tau_s$, so that the superpotential (\ref{eq:Wnonpert}) is dominated by the instanton contribution from branes wrapping the small cycle
\begin{equation} 
	W = W_{0} + A_{s}e^{-c_s T_{s}} \, . \label{eq:Wmin1}
\end{equation}
These assumptions will turn out to be self-consistent.

In our construction we will ultimately consider four stacks of branes. 
The first three wind the large cycles $\tau_1$ and $\tau_2$ multiple times, while the fourth winds the small cycle $\tau_s$ just once, as shown in Table \ref{t:braneconfig}.  We assume that our axions are not related to QCD: an additional D7-brane stack harboring the Standard Model may also be considered separately.

Together, these two types of corrections to $K$ and $W$ conspire to stabilize the K\"{a}hler moduli at parametrically large volume, which we now illustrate for this class of examples.  
The volume (\ref{volume}) provides the K\"ahler potential (\ref{eq:Kpert}),
where $K_0$ contains the contribution from the axio-dilaton and complex structure moduli, which are not relevant to our discussion. 
The induced K\"ahler metric is 
\begin{equation} \label{Kahlerevaluated}
	K_{i\bar \jmath}= \frac{1}{8(\V+\hat \xi/2)^2} 
	\begin{pmatrix}
		\frac{ \alpha \tau_2(2\alpha \tau_1^{1/2} \tau_2 - e)}{\tau_1^{3/2}}  &\frac{2 \alpha   e}{\tau_1^{1/2}} & -\frac{3 \alpha^2 \gamma   \tau_2 \tau_s^{1/2}}{\tau_1^{1/2}} \\
		\frac{2 \alpha   e}{\tau_1^{1/2}}   & 4 \alpha^2 \tau_1 & -6 \alpha^2 \gamma \tau_1^{1/2}  \tau_s^{1/2}  \\
		-\frac{3 \alpha^2 \gamma   \tau_2 \tau_s^{1/2}}{\tau_1^{1/2}}  & -6 \alpha^2 \gamma \tau_1^{1/2}  \tau_s^{1/2} & \frac{3 \alpha \gamma (\alpha \tau_1^{1/2} \tau_2 + 2 e + 3 \hat \xi/2)}{ \tau_s^{1/2}}  \\
	\end{pmatrix}\,,
\end{equation}
where we have defined 
$$e \equiv \alpha \gamma \tau_s^{3/2} - \hat \xi/2\,.$$

Plugging the potentials \eqref{eq:Kpert} and \eqref{eq:Wmin1} into the F-term scalar potential and minimizing yields an LVS solution.  The leading-order scalar potential is
\begin{equation}
	V_{\rm F} \simeq e^{K_{0}}\left[\frac{8\tau_s^{1/2}c_s^{2}A_s^{2}}{3\alpha\gamma\V}e^{-2c_s\tau_s} - \frac{4c_s\tau_sA_s}{\V^{2}}W_{0}e^{-c_s\tau_s} + \frac{3 \hat \xi}{4\V^{3}}W_{0}^{2}\right] \, . \label{VLVS}
\end{equation}
Here we have already minimized with respect to $a_{s}$ by setting 
$$\cos(\theta + c_sa_s) = -1,$$
which generates the minus sign in the second term. We have also allowed for a phase $\theta$ arising in the cross-term potential due to a phase difference between $W_0$ and $A_s$ in (\ref{eq:Wmin1}) \cite{Cicoli:2008va}. Physically this is not a serious problem since the field space of an axion is compact and does not contain runaway directions which could destabilize the vacuum. From \eqref{volume}, \eqref{eq:Kpert} and \eqref{eq:Wmin1} the potential \eqref{VLVS} is a general result to $\mathcal{O}(\V^{-3})$ --- in fact it holds in general ``swiss cheese'' manifolds regardless of how many bulk cycles are present, or their configuration.  Note that it is an approximation at large $\V$: the full result is summarized in eg. equation (17) of Ref. \cite{Balasubramanian:2005zx}.

The potential is asymptotically AdS, approaching zero from below, and has a minimum at \cite{Cicoli:2008gp}
\begin{equation}
	\V \simeq \frac{3\alpha\gamma W_{0}\tau_s^{1/2}e^{c_s\tau_s}}{c_sA_s}\left(\frac{c_s\tau_s-1}{4c_s\tau_s-1}\right) \, , \qquad \hat{\xi} \simeq \frac{32\alpha\gamma c_s \tau_s^{5/2}\left(c_s\tau_s-1\right)}{\left(4c_s\tau_s-1\right)^2} \, . \label{eq:LVSminfibered}
\end{equation}
In the limit $c_s\tau_s\gg 1$ this can be simplified as
\begin{equation}
	\V \simeq \frac{3\alpha\gamma W_{0}\tau_s^{1/2}e^{c_s\tau_s}}{4 c_sA_s} \, , \qquad \tau_s \simeq \left(\frac{\hat \xi}{2\alpha\gamma}\right)^{2/3} \, . \label{eq:LVSminfiberedlargetau_s}
\end{equation}
For weak coupling we see that it is natural to have $\tau_{s} \gtrsim \O(1)$, which leads to an exponentially enhanced volume $\V$.  Thus the expansion in $\V^{-1}$ is controlled, with additional contributions arising at higher order in the expansion, justifying our initial assumptions.\footnote{Certain higher-order corrections may further constrain the allowed range of $\V$, if present \cite{Pedro:2013qga}.} Furthermore, analysis of general manifolds with additional small cycles shows that all such small cycles are stabilized with comparable sizes for a reasonable range of parameters \cite{Cicoli:2008va}.  In this work we will ultimately be steered towards small values of $c_s\tau_s$, for which \eqref{eq:LVSminfiberedlargetau_s} is no longer a good approximation to the full result \eqref{eq:LVSminfibered}.  Nevertheless, as long as $c_s\tau_s\gtrsim 1$, the solution is still well-defined and the resulting modification to \eqref{eq:LVSminfiberedlargetau_s} is by at most an $\mathcal{O}(1)$ factor, so we expect the above argument to still hold.


\subsection{Moduli stabilization II: the other large cycles and uplifting}

At this stage the other large-cycle moduli and the associated axions remain unstabilized. 
We may stabilize them by introducing further corrections to the potential.
Since the overall volume is fixed, further saxion-dependent contributions will fix the relative sizes of the large-cycle volumes.
One conventional approach to this end is to use string loop corrections \cite{Berg:2005ja,Berg:2005yu,Cicoli:2008va,Berg:2007wt,Cicoli:2007xp}.  There are contributions arising from string loops located as follows: within each bulk cycle, corresponding to Kaluza--Klein (KK) modes; or between them, corresponding to winding (W) modes. 
The explicit form is not known for general geometries, but has been calculated for some orbifolds \cite{Berg:2005ja}. For a general Calabi--Yau manifold, the appropriate extension of the orbifold results has been conjectured 
\cite{Berg:2007wt}.

The KK modes arise from the one-loop contribution of open strings between D3 and D7-branes (or O7-planes). These strings can be viewed as closed strings having KK quantum numbers, giving a correction to the K\"ahler potential
\begin{equation}
 \delta K_{g_s}^{\rm KK} \sim \sum_{I=1}^M \frac{m_{I,\rm KK}^{-2}}{({\rm Re}S) \V} \sim g_s \sum_{I=1}^M \frac{{\cal E}_I^{\rm KK}  a_{Ij} t^j}{\V} \, ,
\end{equation}
where ${\cal E}_I^{\rm KK}$ are $\mathcal{O}(1)$ shape coefficients depending on the complex structure. Unless we know the details of the D3-brane configuration, whose details we do not consider but which are necessary for tadpole cancellation, we regard $a_{Ij}t^j$ as the two-cycle volumes transverse to the cycles $D_I = Q_{Ik} D_k$ wrapped by the $I$th D7-brane stack. 

There are also modes coming from winding strings along the intersection between two stacks of D7-branes (or O7-planes), 
\begin{equation} \label{interseccurve}
 C_{IJ} = D_I \cdot D_J = (Q_{Ii} D_i) \cdot (Q_{Jj} D_j) \, .
\end{equation}
Their contribution is 
\begin{equation} \label{windingcorr}
 \delta K_{g_s}^{\rm W} \sim \sum_{I \ne J} \frac{m_{IJ,\rm W}^{-2}}{ \V} \sim \sum_{I \ne J} \frac{{\cal E}_{IJ}^{\rm W}}{a_{IJk} t^k\V} \, ,
\end{equation}
where ${\cal E}_{IJ}^{\rm W}$ are likewise geometric coefficients, and $a_{IJj}t^j$ are the two-cycle volumes of the curves $C_{IJ}$ in (\ref{interseccurve}). 

From the ``extended no-scale structure'' enjoyed by the loop corrections \cite{Cicoli:2007xp}, the resulting correction to the scalar potential is simplified as
 \begin{equation} \label{oneloopcorr}
 \delta V_{g_s} = \left[K^0_{i \bar \imath}  \sum_I  \left(g_s{\cal E}_I^{\rm KK}  a_{Ii}\right)^2  - 2 \sum_{I \ne J} \frac{ {\cal E}_{IJ}^{\rm W}}{a_{IJk} t^k \V} \right] \frac{W_0^2}{\V^2} \, .
\end{equation}
Therefore, we may fix all the bulk-cycle volumes by minimizing the total potential, with the constraint of fixed overall volume (\ref{eq:LVSminfibered}), since it has already been stabilized by the leading contribution \eqref{VLVS}. This is a general expression taking into account generic wrapping and arises at $\mathcal{O}(\V^{-10/3})$.  One may worry that for small volumes, there does not seem to be sufficient parametric suppression relative to the leading LVS potential \eqref{VLVS}, which arises at $\mathcal{O}(\V^{-3})$, in order to ensure the validity of treating \eqref{oneloopcorr} as a small correction.  However, the KK corrections are also suppressed by $g_s^2$, and we are interested in the parameter region where the winding terms give corrections of around the same size.  Thus, simply requiring small $g_s$ seems enough to maintain a robust hierarchy.\footnote{For readers who nevertheless remain skeptical, we point out that our analysis of axion inflation is largely independent of the precise details of moduli stabilization, so the present discussion may be regarded as heuristic.}

First, we calculate the KK contributions.
Let $C_I \equiv d_{Ijk} D_j \cdot D_k$ be the two-cycle transverse to the $I$th brane stack. Its volume is
$$
 a_{Ii} t^i = J \cdot C_I = t^i d_{Ijk} D_i \cdot D_j \cdot D_k = t^i d_{Ijk} \kappa_{ijk} \, ,
$$
thus 
$$
 a_{Ii}  = d_{Ijk} \kappa_{ijk} \, .
$$
Assuming that the relevant branes intersect transversely, we have 
\begin{equation}
 (Q_{Ii} D_i) \cdot C_I =  Q_{Ii}d_{Ijk}  \kappa_{ijk} = Q_{Ii} a_{Ii} = 1 \quad \text{(no sum over $I$)}\, .
\end{equation}
We define $Q_{Ii} a_{Ii}  = h_{Ii}$ for each $i,I$ satisfying $\sum_i h_{Ii} =1$ for all $I$.
Since the detailed locations of the D3-branes are unknown, this constraint is not precise and we use it simply to estimate orders of magnitude.
Thus, we find
$$
 \delta K_{g_s}^{\rm KK} \sim \sum_{i=1,2,s} \frac{g_s}{\V} \sum_{I=1,2,3,s}   \frac{{\cal E}_I^{\rm KK}  h_{Ii}}{Q_{Ii}}t^i \, .
$$
Note that (\ref{oneloopcorr}) does not depend on particular properties of the geometry, such as intersection numbers $\kappa_{ijk}$ or the detailed volumes $a_{Ii}t^i$. We need only the coefficient of each $t^i$ and the D3 configurations fixing $h_{Ii}$. Also we note that, since $\tau_s$ and other small four-cycles do not intersect the bulk cycles, their string-loop corrections do not contribute to moduli stabilization but simply shift the background value of the potential.

Now we calculate the winding contributions.
The intersection curves $C_{IJ}$ in (\ref{interseccurve}) have volumes
\begin{equation}
  J \cdot C_{IJ} = \kappa_{ijk} Q_{Ii} Q_{Jj} t^k\equiv a_{IJk} t^k \, .
\end{equation}
This fixes the volume coefficients in terms of winding numbers,
$$
 a_{IJk} = \kappa_{ijk} Q_{Ii} Q_{Jj} \, ,
$$
where the dependence on the geometry enters as $\kappa_{ijk}$.
Thus, we obtain
$$
 \delta K_{g_s}^{\rm W} \sim
 \sum_{IJ=12,13,23} \frac{{\cal E}_{IJ}^{\rm W}}{\V}\left( \kappa_{ij1} Q_{Ii}Q_{Jj} t^1 + \kappa_{ij2} Q_{Ii}Q_{Jj} t^2  \right)^{-1} \, ,
$$
where $I$ and $J$ run over the brane stacks of Table \ref{t:braneconfig}.  Since there is no intersection between the branes along $D_I$ and $D_s$, there is no corresponding correction. In the alignment limit, the terms with $IJ=13$ and those with $IJ=23$ are proportional to each other.

\begin{figure}[t!]
 \begin{center}
\includegraphics[width=6.5cm]{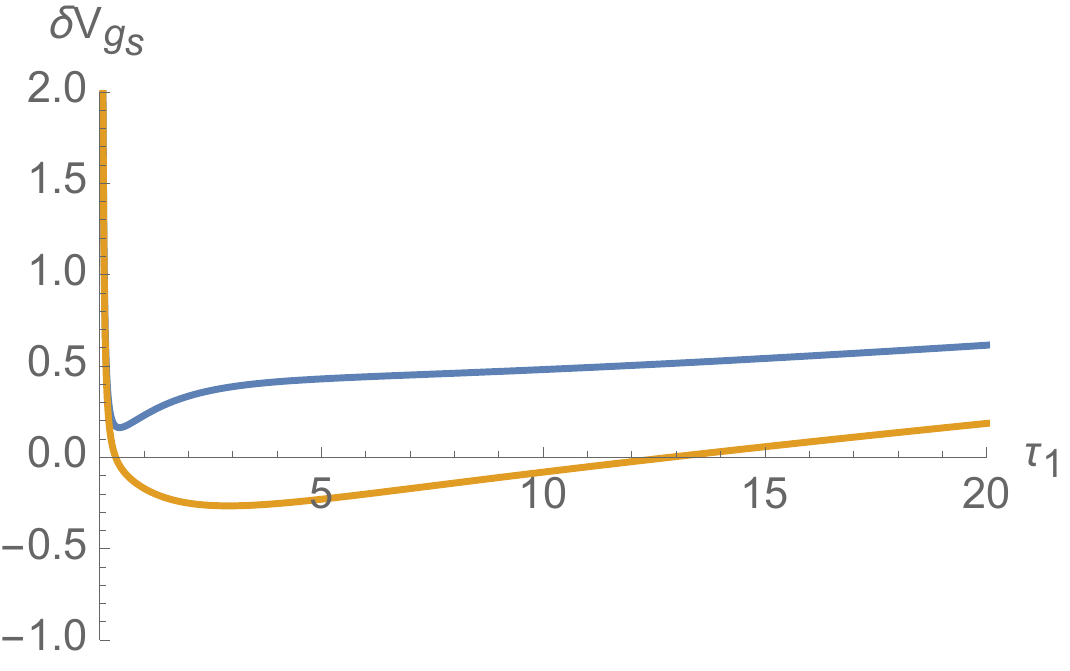} \quad
\includegraphics[width=6.5cm]{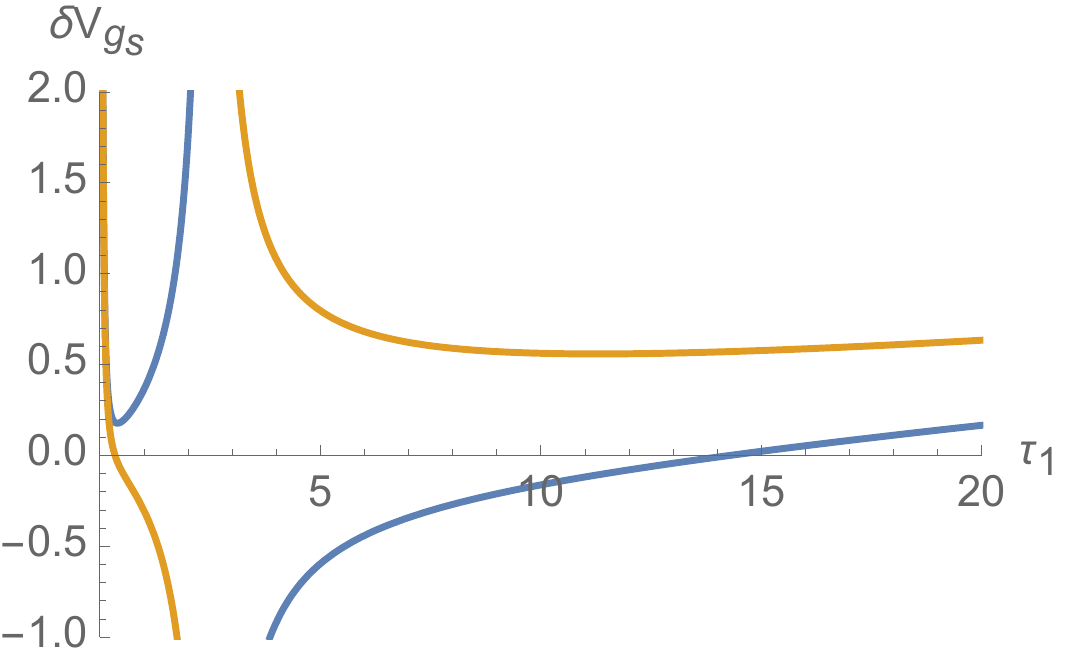} 
\end{center}
\caption{Schematic form of the one-loop potential. The left and right panels correspond to the cases ${\sf cd}>0$ and ${\sf cd}<0$, respectively; within these, the orange curves correspond to ${\sf d} > 0$ while the blue curves correspond to ${\sf d} < 0$. When the signs are the same we have a global minimum; for opposing signs we only have a local minimum. We use $\delta V_{g_s} = (0.1)^2 \tau_1^{-2} + 2 \cdot ( 0.1)^2 \tau_1  \pm ( \sqrt{\tau_1} \pm 4 \tau_1^{-1} )^{-1}.$
} \label{fig:onelooppotential}
\end{figure}
In our reference example ${\mathbb C}\P^4_{[1,1,2,2,6]}(12)$, the only non-vanishing intersection number is $\kappa_{122}=2$, and the two-cycles are $t^1= \tau_2/(2\sqrt{\tau_1}), t^2 = \sqrt{\tau_1}$, as calculated in the Appendix.
The resulting potential is highly nonlinear, due to the volume dependence of the winding contribution. This poses a complication because the expression for the geometric factor ${\cal E}_{IJ}^{\rm W}$ cannot be transformed easily to that of elementary cycles ${\cal E}_{ij}^{\rm W}$, which we would like to define in a similar manner as (\ref{windingcorr}). In light of this, for the purposes of further understanding we analyze moduli stabilization heuristically. We will also neglect the contribution with $IJ=23$.

With the bulk volume (\ref{fiberdvol}), we may reduce one parameter as $\tau_2 = \V \alpha^{-1} \tau_1^{-1/2}$, such that 
\begin{equation}
	\delta V_{g_{s}} = \left[\frac{\sf a}{\tau_{1}^{2}} + \frac{{\sf b} \alpha^2 \tau_1}{\V^2} - \left({\sf c} \alpha\tau_{1}^{1/2}\V + \frac{{\sf d} \V^2}{{\tau_1}} \right)^{-1} \right] \frac{W_{0}^{2}}{\V^2} \, , \label{eq:Vsloop}
\end{equation}
where we define dimensionless parameters
\begin{eqnarray}
	&\displaystyle{\sf a} \equiv\sum_I{\frac{1}{4}}  \left(g_s{\cal E}_I^{\rm KK}  \frac{h_{I1} }{Q_{I1}}\right)^2 \, ,
	 &{\sf b} \equiv \sum_I \frac{1}{2}\left( g_{s}{\cal E}_{2}^{\mathrm{KK}} \frac{h_{I2}}{Q_{I2}} \right)^{2} \, ,\\
	&\displaystyle {\sf c} \equiv \frac{ \kappa_{ij{2}} Q_{1i}Q_{2j} }{{ {\cal E}_{12}^{\mathrm{W}}}} \, ,
	 & {\sf d}\equiv \frac{ \kappa_{ij{1}} Q_{1i}Q_{2j} }{2 {\cal E}_{12}^{\mathrm{W}}} \, .
\end{eqnarray}
We note that ${\sf a}>0$ and ${\sf b}>0$, and these terms are suppressed by $g_s^2$. We can see that for ${\sf cd}>0$, the scalar potential has a global minimum. Interestingly, for ${\sf cd}<0$, the potential has only a local minimum. The schematic form is shown in Fig. \ref{fig:onelooppotential}. Depending on the geometry, 
the minimum is formed at one of the following compromising points: 
\begin{equation} \label{stabilizedtaus}
 \frac{\sf a}{\tau_1^2} \sim \frac{1}{{\sf c} \alpha\tau_1^{1/2} \V} \, , \qquad 
  \frac{\sf a}{\tau_1^2} \sim \frac{\tau_1}{{\sf d} \V^2} \, , \qquad 
 \frac {{\sf b}{\alpha^2}\tau_1}{\V^2} \sim \frac{1}{{\sf c} {\alpha}\tau_1^{1/2} \V} \, , \qquad 
 \frac {{\sf b}{\alpha^2}\tau_1}{\V^2} \sim \frac{{\tau_1}}{{\sf d} \V^2} \, .
\end{equation}
All of these imply that stabilization occurs around 
\begin{equation}
\frac{\tau_2}{\tau_1} \sim \frac{{\cal E}_{12}^{\mathrm{W}}}{g_s^2 ({\cal E}_I^{\rm KK})^2} \, . 
\end{equation}
For $\mathcal{O}(1)$ values of ${\cal E}_{12}^{\mathrm{W}}$ and ${\cal E}_I^{\rm KK}$, this fraction can be close to one since $g_s<1$.

In order to make contact with our observed universe we must uplift the AdS vacuum to Minkowski or de Sitter.  There are various mechanisms to realize this, for example by introducing anti-D3 branes \cite{Crino:2020qwk} (see \cite{Cicoli:2012fh,Cicoli:2015ylx,Gallego:2017dvd} for other possibilities).  
For the case of anti-D3 branes, in order not to destabilize the entire construction, the anti-branes must be sequestered to a different region of the geometry, for example a highly throat.  In the four-dimensional effective description, this induces an uplifting potential at $\mathcal{O}(\V^{-2})$.  
For sufficiently large volume, the shift of the vacuum induced by uplifting is negligible,\footnote{For a simple justification, see Appendix A of \cite{Cribiori:2021gbf}.} so we may to a good approximation use the values of the moduli obtained in the LVS mechanism.

However, as is well known, consistency with the stringent phenomenological bounds on the vacuum energy of the universe requires a high degree of tuning, potentially of order $10^{120}$.  In the present scenario, this tuning is divided into two steps.  First of all, the potential after moduli stabilization and uplifting should yield a small positive constant at the scale of the light-axion potential (to be presented in the next subsection).  Furthermore, the minimum of that potential should approach zero; in particular, it should be tuned to the observed cosmological constant.  It would be interesting to explore whether at least the first step could be achieved via a correlation with the exponentially-suppressed light-axion potential. 

\subsection{Restricted parameters} \label{sec:restrictedparams}

So far we have stabilized all the K\"ahler moduli and the axions associated with the small cycles.  Only the bulk axions remain unfixed: they can only receive masses from non-perturbative corrections due to branes wrapping bulk cycles, leading to masses which are exponentially suppressed by bulk volume moduli.  Traditionally the Large Volume Scenario has been used to obtain light superpartners at the TeV scale, in order to solve the electroweak hierarchy problem.  In such cases the bulk-axion masses become vanishingly light, and those axions behave as dark radiation.  On the other hand, if we set aside this motivation and instead consider small volumes $\V$, the axions may remain sufficiently heavy that
they can instead be used to achieve natural inflation via the alignment mechanism.

We continue to analyze the vacua from the above fibered geometry with two axions.
Having stabilized all other moduli, 
the leading K\"ahler metric for the bulk axions becomes
\begin{equation} \label{twobytwoKahler}
 \frac{\hat{K}_{i j}}{M_{\rm Pl}^2} = 
   \begin{pmatrix}
   \frac{1}{ {4}\tau_1^{2}}  & 0  \\
0 & \frac{1}{{2}\tau_2^{2}} \\
   \end{pmatrix} \, ,
\end{equation}
where we have approximated $\V \gg \hat \xi$. In our case, the K\"ahler metric becomes approximately diagonal due to the structure (\ref{fiberdvol}) and moduli stabilization of small cycles (\ref{eq:LVSminfibered}), which from \eqref{Kahlerevaluated} implies that $e \sim \alpha\gamma\tau_s^{1/2}/c_s \ll \V$. Recall that the K\"ahler cone condition restricts the allowed range of the moduli: for example, in our reference geometry ${\mathbb C}\P^4_{[1,1,2,2,6]}(12)$, the condition (\ref{ourkahlercone}) constrains $\tau_2 > \frac43 \tau_1$.
With the normalization as in (\ref{kinetic}), the axion decay constants are
\begin{equation} \label{decayconsts}
 f_1 = \frac{M_{\rm Pl}}{2 \sqrt 2 \pi \tau_1} \, , \qquad f_2 = \frac{M_{\rm Pl}}{2 \pi \tau_2} \, .
\end{equation}

We assume that the leading scalar potential is stabilized around the uplifted LVS minimum, with its dependence on the other moduli integrated out and represented as a small, positive constant $V_{0}$.  We also assume that gaugino condensation on the D7-brane stacks listed in Table \ref{t:braneconfig} give rise to a non-perturbative superpotential of the form \eqref{eq:Wnonpert}.  Then the leading-order low-energy effective potential for the bulk axions takes the form\footnote{{Note that a bulk-axion dependence may also arise as a non-perturbative correction to the K\"ahler potential (we thank Joe Conlon for pointing this out).}  Fortunately, it turns out that such corrections enter at subleading order in $\V$ compared to \eqref{finalpotential}, thanks to the no-scale structure.}
\begin{equation} \label{finalpotential}
 V = V_0 - \lambda_1^4 e^{-S_1} \cos(c_1 Q_{1j}b_j) - \lambda_2^4 e^{-S_2} \cos(c_2 Q_{2j} b_j) -  \lambda_3^4 e^{-S_3} \cos(c_2 Q_{3j}b_j) \, ,
\end{equation}
with coefficients
$$
 \lambda_I^4 = \frac{4 e^{K_0} W_0}{\V^2} A_I S_I ,
$$
where the intstanton actions $S_I$ are defined as in (\ref{instaction}). As discussed in the previous subsection, the constant $V_{0}$ should be chosen such that at the scale of inflation, the minimum of the potential is approximately zero.

The potential \eqref{finalpotential} is of the form (\ref{axionLattice}), which may realize KNP natural inflation via enhancement of the effective decay constant due to alignment (\ref{alignment}).  As discussed in section \ref{sec:tunneling}, another potential source of enhancement is the group theory factor $N_I$ from the gauge condensate, reflected in $c_I = 2\pi / N_I$. 


The K\"ahler moduli are subject to the CHC of the weak gravity conjecture.
Since the third potential is introduced to remedy the CHC, we want to ensure that its dynamics do not affect those of the other two instanton contributions and spoil axion alignment. This implies that its action should be sufficiently large,
\begin{equation} \label{thirdinst}
 S_3 =\frac{2\pi}{N_3} ( Q_{31} \tau_1 + Q_{32} \tau_2) \gg 1 \, ,
\end{equation}
to suppress its contribution to the potential. 
Since $\tau_1,\tau_2$ are positive definite, we may choose 
\begin{equation} \label{thirdcharges}
 Q_{31}>0 \, , \qquad Q_{32} >0 \, ,
\end{equation}
whose typical ${\cal O}(1)$ values are enough to suppress the potential. It is safe to have small values of $N_3$, so we take it to be 1.

\begin{figure}[t!]
 \begin{center}
\includegraphics[width=8cm]{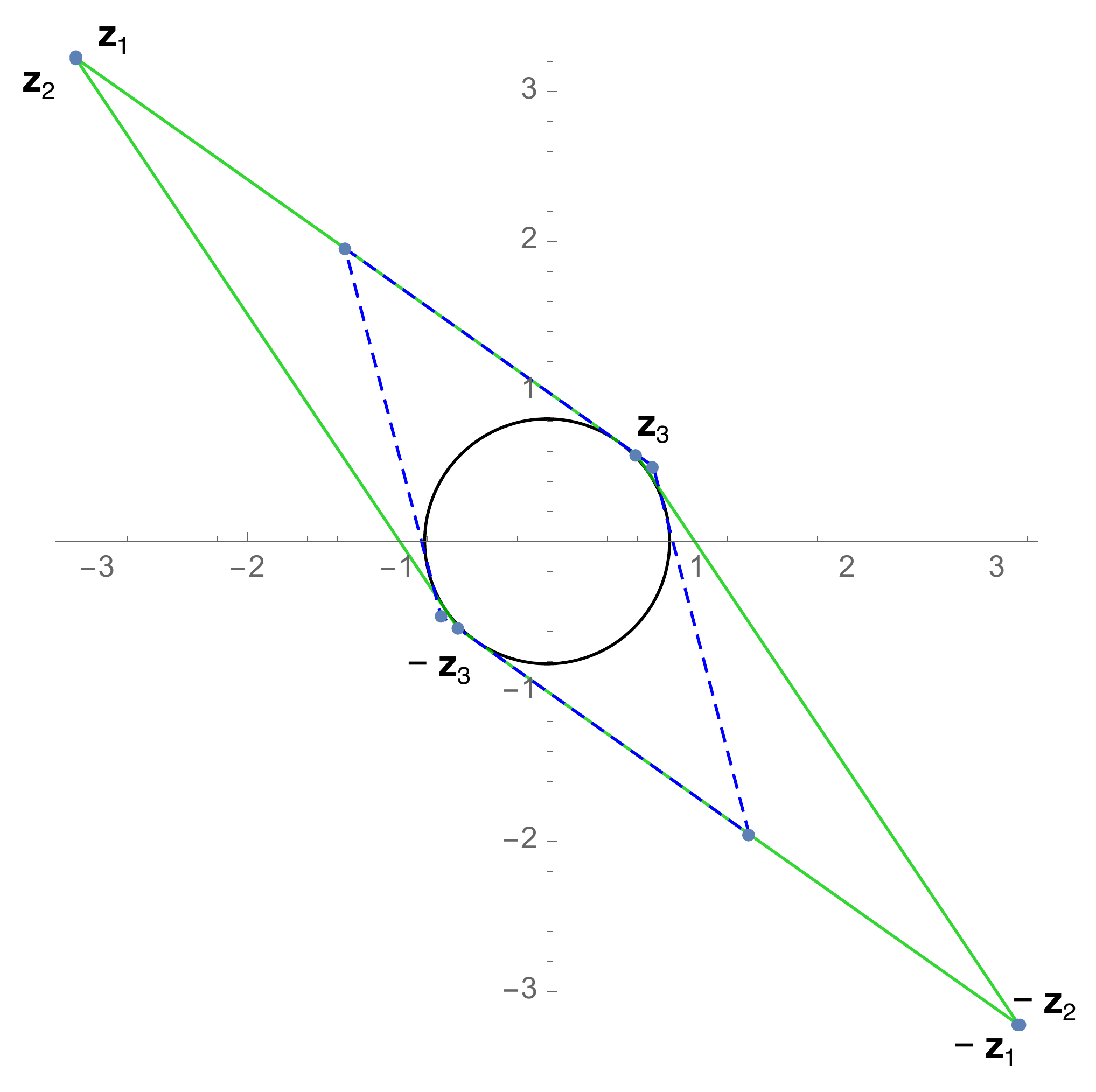}
\end{center}
\caption{The convex hull condition (CHC) from the weak gravity conjecture. 
The convex hull spanned by all the normalized charge vectors ${\bf z}_I$ should include the circle of radius $\r$ defined in (\ref{ctomass}). We place the vector ${\bf z}_3$ in the first quadrant in order to obtain a large action and hence suppress the corresponding instanton contribution. In order to satisfy the CHC, the original lattice vectors ${\bf z}_1, {\bf z}_2$ should then lie in the second or fourth quadrant. The solid and dashed convex hulls correspond to the cases $Q_{I2} \tau_2 = \sqrt{2} Q_{I1} \tau_1$ and $Q_{I2} \tau_2= 2 Q_{I1}  \tau_1$, respectively. Note that all ${\bf z}_I$ vectors for the original instantons lie on the same line. 
The CHC favors isotropic geometry for a uniform choice of charges. 
} \label{fig:CHC}
\end{figure}

For the instanton actions (\ref{instaction}), the charge-to-mass ratio vectors (\ref{zvector}) are
\begin{equation}  \label{thirdWGCvec}
	\mathbf{z}_I = \left(\begin{array}{c} \sqrt{2} \frac{Q_{I1}\tau_1}{Q_{I 1}\tau_1 + Q_{I 2}\tau_2} \\ 
		\frac{Q_{I2} \tau_2}{Q_{I1}\tau_1+Q_{I2}\tau_2} \end{array}\right) 
	= \left(\begin{array}{c} \sqrt{2} u_I \\   1-u_I\end{array}\right).
\end{equation}
Note that the enhancement factors of $N_I$ appear in both the instanton actions and the effective decay constants, so taken together, the charge-to-mass ratio vectors are independent of $N_I$.  In order to satisfy the near-orthogonality condition (\ref{CHCorthogonality}), the pairs of charges in the other potentials should have opposing signs,
\begin{equation} \label{oppositesigns}
 Q_{11}Q_{12} <0 \, , \qquad Q_{21}Q_{22}<0 \, .
\end{equation}
This condition places the ${\bf z}_1,{\bf z}_2$ vectors in either the second or the fourth quadrant. In the $\mathbb{CP}^4_{[1,1,2,2,6]}(12)$ model, the K\"ahler condition requires that $\tau_2>\tau_1$, and both vectors should lie in the second quadrant,\footnote{Here we assume that the charges have similar magnitudes.}
\begin{equation} \label{secondquadrant}
 Q_{11} <0 \, , \qquad Q_{12}>0 \, , \qquad Q_{21}<0 \, , \qquad Q_{22}>0 \, .
\end{equation}
This naturally makes the actions $S_1,S_2$ small such that the axions become relatively heavy, as we shall see in Section \ref{sec:observables}. In order to have a valid instanton, its action must be positive, $S_I>0$, which limits us to the range  \begin{align*}
|Q_{22}|\tau_2 > |Q_{21}|\tau_1 \, , \qquad |Q_{12}|\tau_2 > |Q_{11}|\tau_1 \, .
\end{align*} 
Recall that the signs of the charges are determined by the winding numbers of D-branes (\ref{fourcycles}). Negative charges $Q_{Ij}$ are allowed, as long as the resulting cycle $D_I$ is effective.

Considering positive charges, one can identify several interesting limits: 
\begin{equation} \label{znorm}
\begin{split}
\mathbf{z}_I &\to \left(\begin{array}{c} \sqrt{2} \\ 0 \end{array}\right) \quad {\rm for}\ Q_{I1} \tau_1 \gg Q_{I2} \tau_2
\ \Rightarrow |\mathbf{z}_I | \to \sqrt{2} \, ;
\nonumber\\
&\to \left(\begin{array}{c} 0  \\ 1  \end{array}\right) \quad\;\;\; {\rm for}\ Q_{I2}  \tau_2 \gg Q_{I1} \tau_1\ 
\Rightarrow |\mathbf{z}_I| \to 1 \, ; \nonumber\\
&\to  \left(\begin{array}{c}  \frac{\sqrt{2}}{3} \\ \frac{\sqrt{2}}{3}  \end{array}\right) \;\;\,\,\, {\rm for} \ \frac{Q_{I1}\tau_1}{Q_{I2} \tau_2} = \frac{1}{2}
\ \Rightarrow |\mathbf{z}_I| \to  |\mathbf{z}_I|_{\rm min}=\sqrt{\frac{2}{3}} \, .
\end{split}
\end{equation} 
In the plane of $\mathbf{z}_I = (x_I, y_I)$, the charge-to-mass ratio vectors $\mathbf{z}_I$ for these string axions span the line
\begin{equation} \label{zline}
\frac{x_I}{\sqrt{2}} + y_I = 1 \, .
\end{equation} 
For instantons with net positive charge, $S_I > 0,I=1,\dots,M$, all ${\bf z}_I$ vectors lie on the same line (\ref{zline}). We observe that this line is tangent to the circle of radius $\r = \sqrt{2/3}$. Therefore, the part of the convex hull generated by the corresponding ${\bf z}_I$ vectors always contains the circle, by construction.

As we have seen above, the nontriviality arises from the third potential in (\ref{finalpotential}), which is introduced to solve the CHC, which corresponds to a charge-to-mass ratio vector 
\begin{equation} 
\mathbf{z}_3 = \left(\begin{array}{cc} \sqrt{2} u_3 \\ 1- u_3 \end{array}\right) \quad (0<u_3 <1)
\end{equation} 
lying in the first quadrant, from (\ref{thirdcharges}).
The first and second instantons have $u_1 <0$ $(1-u_1>0)$, $u_2 <0$ $(1-u_2 >0)$, so they are located in the second quadrant. 
The reason the choice of ${\bf z}_3$ is nontrivial comes from the contribution of the anti-instantons, which have opposite charges but the same action, $S_I = c_I |Q_{Ii}\tau_i|, I=1,\dots,M$ for $Q_{Ii}\tau_i < 0$. 
The convex hull formed by an instanton and anti-instanton do not trivially contain the circle. In our example, we need to impose one further condition: namely, the portion of the convex hull spanned by $\mathbf{z}_3$ and $-\mathbf{z}_I$, for $I=1,2$, should contain the circle of radius $\r$.  Thus, we require
\begin{equation} \label{CHCinequality}
u_3 \geq \frac{1-u_{I}}{1-3u_{I}} \ \Rightarrow 
 \frac{Q_{31}}{Q_{32}}  \geq  \left(-\frac{Q_{{I}2}\tau_2}{Q_{{I}1}\tau_1}\right)\left(\frac{\tau_2}{2\tau_1}\right) \, .
\end{equation} 
Given charges of two instantons, $(Q_{11},Q_{12},Q_{21},Q_{22})$, we may always find charges of the third instanton $(Q_{31},Q_{32})$ satisfying this relation by taking large enough $Q_{31}/Q_{32}$. 
This guarantees that the convex hull spanned by all the vectors $\pm \mathbf{z}_I,I=1,2,3$, contains the circle. 

Conversely, by fixing all the charges we can interpret \eqref{CHCinequality} as a limit of the ratio of the K\"ahler moduli,
\begin{equation} \label{nearlyisotropic}
 \frac{\tau_2^2}{2\tau_1^2} \le \frac{Q_{31} |Q_{11}|}{Q_{32}|Q_{12}|} \, .
\end{equation}
Interestingly, the K\"ahler cone condition imposes a limit in the opposite direction.  For example, in ${\mathbb C}\P^4_{[1,1,2,2,6]}(12)$ the condition (\ref{ourkahlercone}) implies that
\begin{equation} \label{nearlyisotropic2}
 \frac{8}{9} \le \frac{\tau_2^2}{2\tau_1^2} \, .
\end{equation}
We note that this condition is model-dependent and it is not clear for general geometries whether the K\"ahler condition works in this way. For two axions with an arbitrary K\"ahler potential, the above relation can be generalized to
\begin{equation}
 \frac{f_1^2}{f_2^2} = \frac{p_1 \tau_2^2}{p_2 \tau_1^2} \, ,
\end{equation}
where $f_1^2/2,f_2^2/2$ are the eigenvalues of the K\"aher metric and $p_i$ are the power-dependences of the moduli $\tau_i$ in the volume (\ref{volansatz}).
We will investigate this relation further in Section \ref{sec:observables}.

The resulting convex hull is illustrated in Fig. \ref{fig:CHC}. The solid and dashed convex hulls correspond to the cases $Q_{I2} \tau_2 = \sqrt{2} Q_{I1} \tau_1$ and $Q_{I2} \tau_2= 2 Q_{I1}  \tau_1$, respectively. As discussed above, the vectors do not depend on the absolute sizes of the $\tau_i$s but only the ratios between them. 
Finally, we note that introducing further brane stacks charged under $b_1$ and $b_2$ does not modify the CHC.
Once we have at least two instantons satisfying the CHC, further inclusion of {\em any} number of instantons only widens the convex hull and thus the CHC is still satisfied.

\subsection{Statistics of charge distribution}

\begin{figure}[t!]
 \begin{center}
\includegraphics[width=6.5cm]{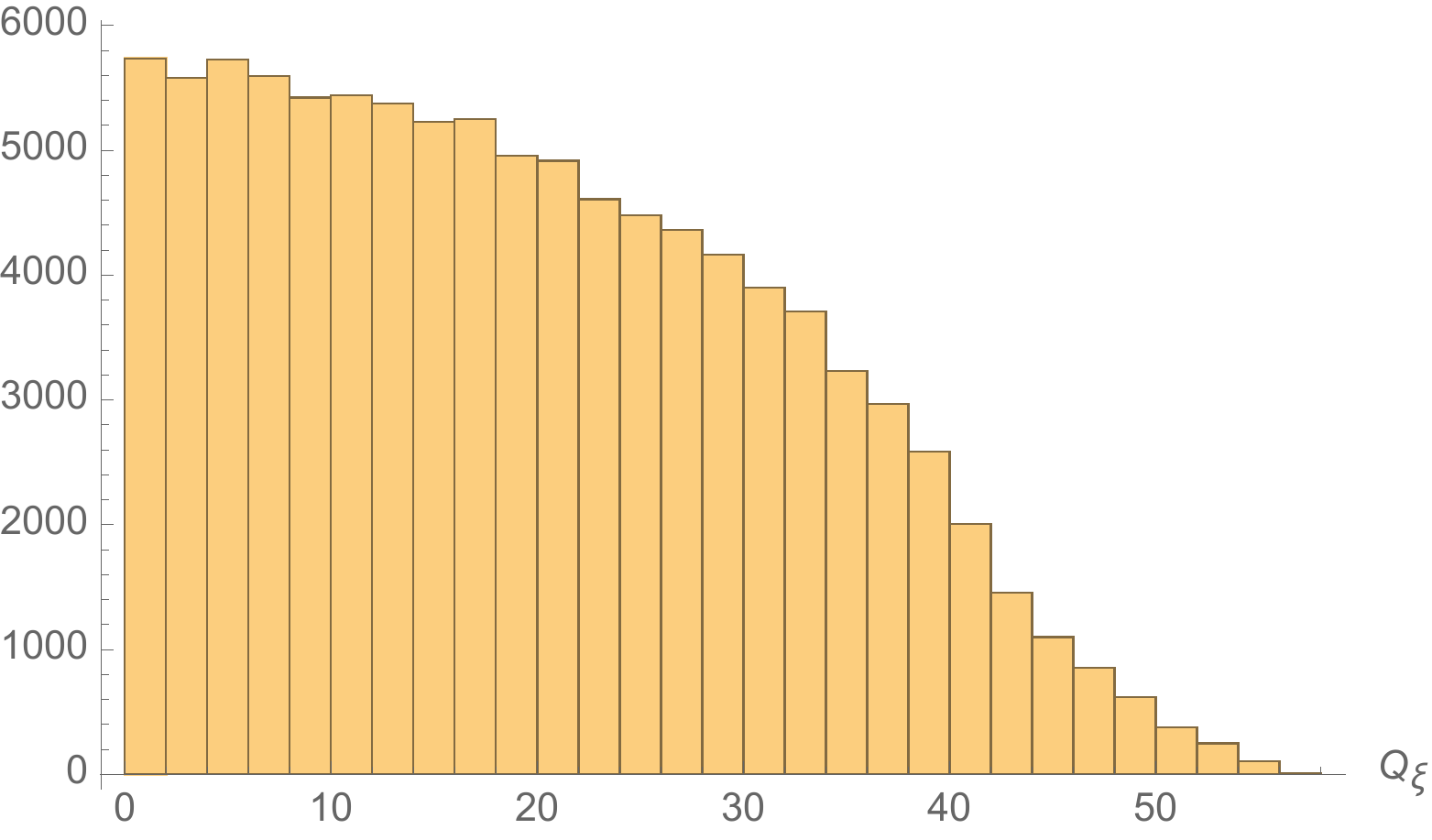}
\includegraphics[width=6.5cm]{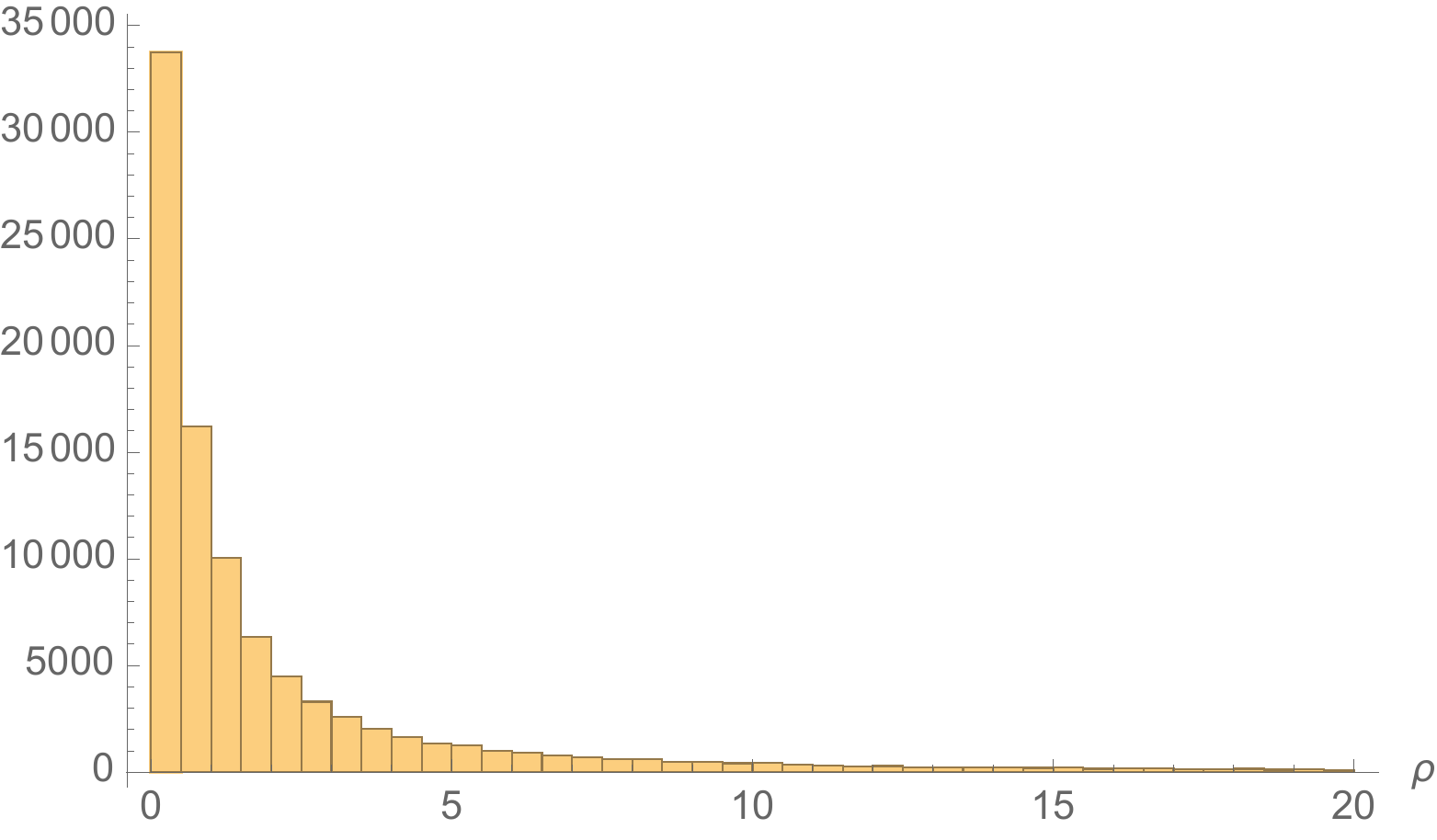}
\end{center}
\caption{  {\em Left:}
Distribution of the alignment $Q_\xi$ defined in (\ref{effcharge}). We allowed random distribution of integers $Q_{Ij}$, $I=1,2,3$, $j=1,2$, from $-40$ to $40$. We need high degree of alignment, or small $Q_\xi$. {\em Right:} Distribution of $\rho$ defined in (\ref{rhoparam}), which should be greater than the ratio of decay constants (\ref{nearlyisotropic}). We see that $\rho\sim 1$ is strongly favored. } \label{fig:distribalign}
\end{figure}

At present, there is no dynamical explanation of brane configurations. Even if we take a top-down approach, we must assemble the D7-brane configuration by hand. Consequently, one may question how likely it is to obtain alignment enhancing the axion decay constant and simultaneously have the third instanton to cure the CHC. 

In this section, we briefly try to understand the behavior via statistical analysis. 
To this end, we generate 100,000 sextuples, $Q_{Ij},I=1,2,3,j=1,2$, of instanton charges randomly, in which each component follows a uniform distribution.

First of all, we consider the alignment condition of two axions. It is characterized by two pairs of lattice vectors, $(Q_{11},Q_{12})$ and $(Q_{21},Q_{22})$. We chose $Q_\xi$ defined in (\ref{effcharge}) as a measure of the (mis)alignment, as discussed in Section \ref{sec:alignment}. 
As the left panel in Fig. \ref{fig:distribalign} shows, there are around 2932 tuples with small enough $Q_\xi$, or large alignment,
\begin{equation}
 P( Q_\xi \le 1) \simeq 2.9\%,
\end{equation}
which shows that the alignment is still probable.

Now we include the effect of the third instanton and axion, which is required to solve the CHC.
If the bulk moduli fields are all stabilized, then the relation (\ref{nearlyisotropic}) gives a restriction on the ratio $Q_{31}/Q_{32}$. Conversely, if we assign the charges $(Q_{31},Q_{32})$, then it is equivalent to setting an upper bound on the ratio $\tau_2/\tau_1$. Thus, we consider the ratio
\begin{equation} \label{rhoparam}
 \rho \equiv \left| \frac{Q_{31} Q_{11}}{Q_{32}Q_{12}} \right| \, .
\end{equation}
The distribution is shown in the right panel of Fig. \ref{fig:distribalign}.
More than 50312 tuples lie in the range $\rho \le 1$,
\begin{equation}
  P( Q_\xi \le 1) \simeq 50.3\% \, .
\end{equation}

Interestingly, even if we apply the K\"ahler cone condition, we still have as many as 3125 tuples,
\begin{equation}
  P \left( \frac89 \le Q_\xi \le 1 \right) \simeq 3.1\% \, .
\end{equation}
Thus, if we choose $Q_{Ij}$s randomly, there is a sizable probability to have a bound close to $\rho = 1$. This should be, because we randomly generated $Q_{Ij}$ with uniform distribution, the inverse ratio $1/\rho$ should follow the same distribution. However, if the one-loop moduli stabilization discussed in 4.2 chooses a large ratio $\tau_2^2/(2\tau_1^2)$, the configuration satisfying this shall be rare.

Combining these two conditions, we find 1514 multiples satisfying $Q_{\xi} \leq 1$ and $\rho \le 1$,
\begin{equation}
 P \left(Q_{\xi} < 1 , \rho \le 1 \right) \simeq 1.5\%,
\end{equation}
which is not small in the string landscape. The result is shown in the left panel of Fig. \ref{fig:distribalign}.
In string theory, however, the axion charge distribution is not uniform. Since D-branes wrap compact cycles, they are subject to the RR tadpole cancellation condition \cite{Brunner:1999jq,Blumenhagen:2002wn}. Roughly, the homology (K-theory) sum of the charges of D-branes $N_IQ_{Ij}$ and their mirror branes $N_IQ_{Ij}'$ with respect to the orientififold planes should be cancelled by the orientifold charge $Q_{Oj}$,
$$
 \sum_{I=1}^M N_I (D_I + D_I') -8 D_{\rm O} =  \sum_{I=1}^M \sum_{j=1}^{h^{1,1}}(N_I Q_{Ij} + N_I Q_{Ij}' - 8 Q_{Oj} ) \tau_j =0 \, ,
$$
where $D_I$ also includes the stack of Standard Model branes (which we have not discussed in this work, but should also be present). The orientifold plane is an invariant plane under an involution $\sigma$, $\sigma D_{\rm O} = D_{\rm O} \equiv Q_{Oj} D_j$, which is an inversion of one coordinate in our construction; see Appendix. This O7-plane carries RR charge $-8$ once we also include those of the image branes.  The image-brane stack is supported by a hypersurface transformed by this involution, $\sigma D_i = D_I'$.

This implies that (i) we inevitably have negative charges, 
$$Q_{Ij}'|_{\rm transv.} = - Q_{Ij}|_{\rm transv.} <0 \, ,$$
transverse to the orientifold plane, as required in (\ref{oppositesigns}), and (ii) there may be an upper limit of the allowed number of D-branes. On a smooth Calabi--Yau manifold the maximal allowed gauge group is $SO(32) \times Sp(16)$ (where $Sp(1)$ is isomorphic to $SU(2)$).   This implies that the charges cannot be comparably larger than 64, giving an upper limit on the allowed charge \cite{Berkooz:1996dw,Choi:2006th,Gmeiner:2005vz}. However, if singularities are present, larger-rank gauge groups may be allowed \cite{Aspinwall:1997ye,Choi:2017vtd}. The tadpole cancellation condition also implies that, in the presence of supersymmetry, the brane configuration can be obtained from deformation, and thus the distribution should not be uniform. A full analysis requires the embedding of the Standard Model, so instead we take a bottom-up approach and approximate the distribution of D-branes as uniform.

\subsection{Small misalignment and observables} \label{sec:observables}

\begin{figure}[t]
	\begin{center}
		\includegraphics[width=12cm]{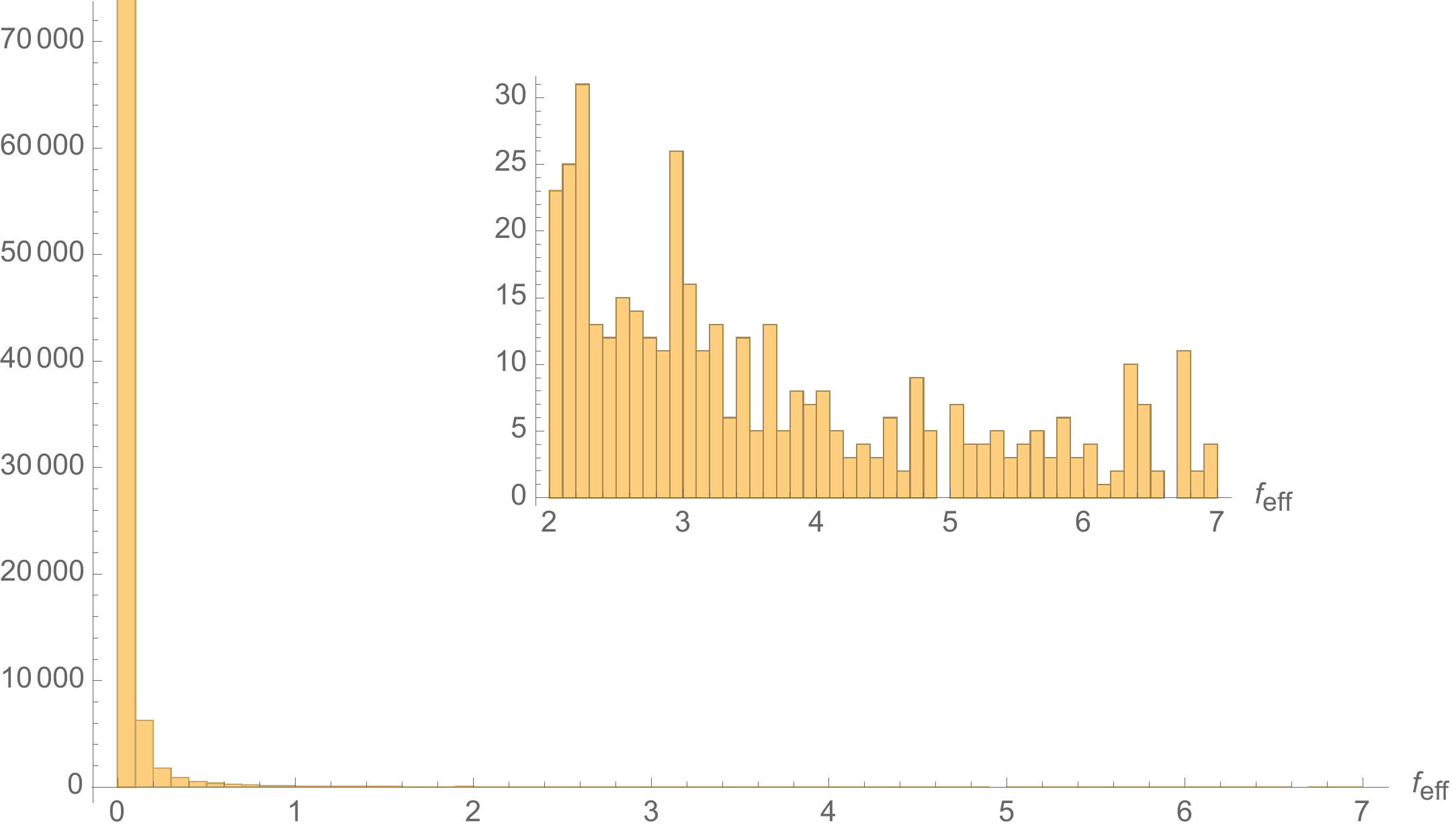}
	\end{center}
	\caption{Distribution of effective decay constants in units of the Planck mass, from 100,000 samples. We assume a uniform distribution of quadruples $(Q_{11} , Q_{12} , Q_{21} , Q_{22})$ as above. We take the volume and parameters as in the second row of Table \ref{t:values}. The inner panel provides a zoomed-in view of the range $2 \le f_{\rm eff} \le 7$. \label{fig:feffpopulation}}
\end{figure}

A small departure from perfect alignment gives one light axion, corresponding to the axion $\phi_\xi$ in Section \ref{sec:alignment}. It may have an effectively larger decay constant \cite{Kim:2004rp}, as is necessary for inflation.
The mass and decay constant of the light axion are \cite{Kappl:2014lra}
\begin{equation}
\begin{split}
 m_\xi^2 &\simeq \frac{\Lambda_2^4}{f_{\rm eff}^2} \, , \\
  \Lambda_2^4 &= \frac{4 e^{K_0} W_0}{ \V^2}  A_2 S_2 e^{-S_2} \, ,
\\
  f^2_{\rm eff} &= \frac{f_1^2 Q_{12}^2+f_2^2 Q_{11}^2}{(Q_{21}Q_{12}-Q_{11}Q_{22})^2} \, .
  \end{split}
\end{equation}
We plot the distribution of effective axion decay constants arising as a function of randomly generated charges $(Q_{11} , Q_{12} , Q_{21} , Q_{22})$ ranging from $-40$ to $40$, as before, in Fig. \ref{fig:feffpopulation}.
Although smaller $\tau_i$s generally lead to larger effective decay constants, the distribution of actual occurrences of large $f_{\text{eff}}$ is insensitive to the specific values of the $\tau_i$s.

In what follows, we take a number of benchmark points in order to check the naturalness of the parameters and to compare with observations.
From the LVS mechanism, the volume $\V$ and the small cycle $\tau_s$ are stabilized as in (\ref{eq:LVSminfibered}). We take $\alpha=\frac12$, $\gamma=1$, $K_0=W_0=1$ and $\chi(X)=-100$, in order to have $\hat \xi = 7.66 g_s^{-3/2}$ and
\begin{equation} 
\tau_s \simeq 0.3887/g_s\, , \qquad \V \sim e^{2\pi \tau_s/N_s} \, .
\end{equation}
Although the string coupling $g_s = {\rm Re} S$ is fixed by fluxes along the lines of GKP \cite{Giddings:2001yu}, as discussed below Eq. (\ref{WGKP}), here we consider it as a free parameter. We are also free to choose $N_s$.

Furthermore, the height of the axion potential should be consistent with the observation (\ref{heightpotential}),
\begin{equation} 
 \frac{\Lambda_2^4}{M_{\rm Pl}^4} \sim e^{-2\pi (Q_{22} \tau_2 + Q_{21}\tau_1)/N_2} \sim 10^{-9} \, .
\end{equation}
This also requires some degree of isotropy, $(Q_{22} \tau_2 - |Q_{21}|\tau_1)/N_2 \simeq 3.3$, which is consistent with the saturated values of the CHC in (\ref{nearlyisotropic}),
\begin{equation} \label{Pconstraint}
 \frac{\tau_2}{\tau_1} \sim \left| \frac{Q_{21}}{Q_{22}} \right| \sim   \left| \frac{Q_{11}}{Q_{12}} \right| \, .
\end{equation}
This is to be compared with the inequalities from both the CHC (\ref{nearlyisotropic}) and the K\"ahler cone condition (\ref{nearlyisotropic2}).

\begin{table}
\begin{center} 
\begin{tabular}{ccccccccc} \hline \hline
 $g_s$ & $N_s$ & $\tau_s$ &  $\V$ & $\tau_2/\tau_1$ &  $\tau_1$ &  $\tau_2$ & $N_1$ & $N_2$ \\  \hline
  0.1 & 7 & 4.336 &33.86 & $\sqrt{2}$ &13.187 & 17.649 & 18 &16 \\
  0.1 & 10 & 4.577 & 16.186 & $\sqrt{2}$ & 8.06 & 11.40 & 16 &14\\ 
  0.05 & 20 & 9.1547 & 45.781 & $4/3$ & 16.77 & 22.36 & 22 &20 \\ 
  \hline
\end{tabular}
\end{center}
\caption{Some benchmark values of moduli fields. We also take $\alpha=\frac12,\gamma=1,K_0=W_0=A_s =1$.}
\label{t:values}
\end{table}
Further taking $A_s =1$, the overall volume and bulk four-cycle volumes are listed in Table \ref{t:values}. All parameters are measured in units of the string length. We may arrange a large hierarchy between the two axion potentials $\Lambda_1 \gg \Lambda_2$ by choosing either $N_1>N_2$ or smaller charges $(Q_{11},Q_{12})$ than $(Q_{21},Q_{22})$.  Moreover, the volume itself should be large enough to ensure validity of the Large Volume Scenario, such that all the various steps in the construction --- the leading mechanism relying on $\alpha^{\prime 3}$-corrections, the stability of further one-loop corrections, even the supergravity approximation itself --- can be trusted as a result of sufficient hierarchy and scale separation.  In this regard, the extended no-scale structure in LVS turns out be a crucial benefactor, fortifying the scenario due to intrinsic suppression of various corrections.  Thus, we may consistently take relatively small volumes, $\V \sim \mathcal{O}(10^{1\text{--}2})$: this is the not-so-Large Volume Scenario, or large volume scenario (lvs).

\begin{figure}[t]
 \begin{center}
\includegraphics[width=7cm]{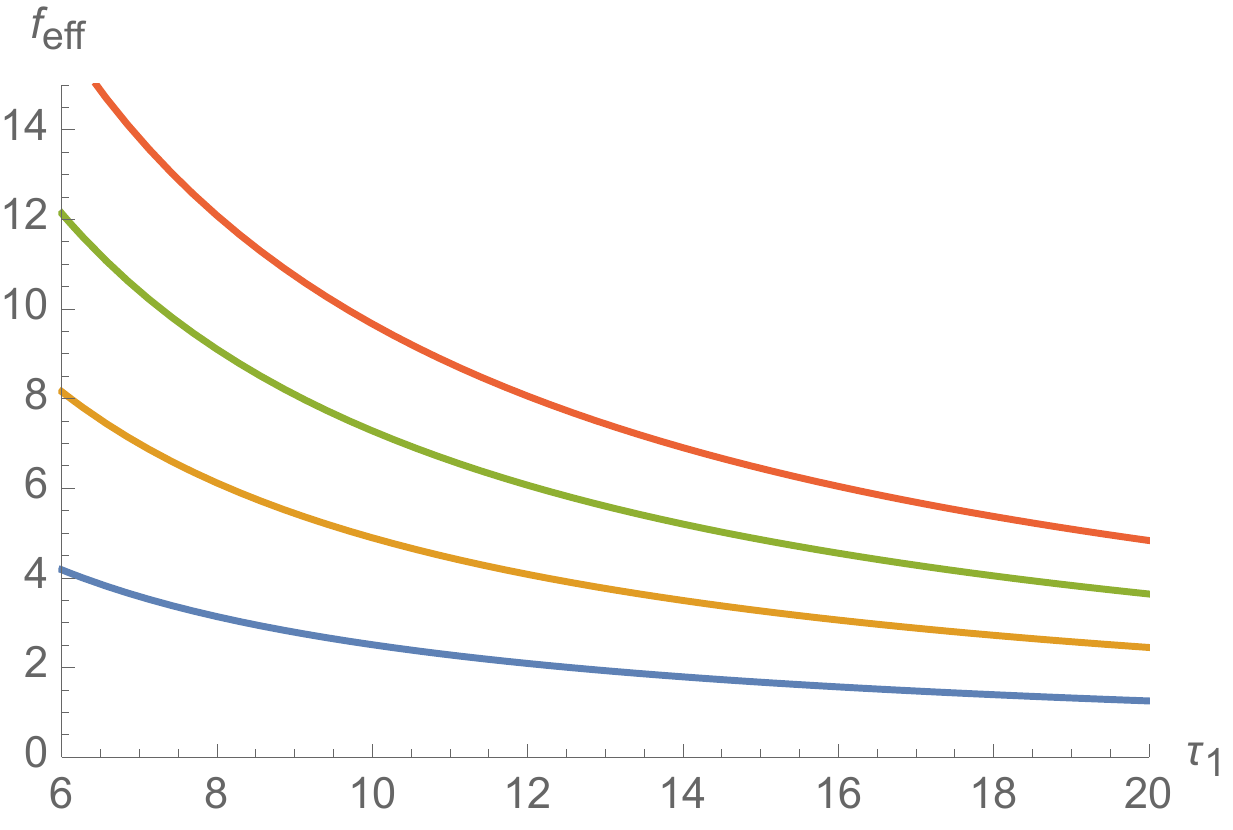}
\end{center}
\caption{Effective decay constant in the unit of Planck mass, as a function of $ \tau_1 = \tau_2/\sqrt{2}$, in the limit $Q=10,20,30,40$ (respectively from below), using the charge ansatz (\ref{alignansatz}). We take $N_1=16,N_2=14$.  }
\label{fig:feff}
\end{figure}

Since the charges are integrally quantized, it is useful to parameterize them using a single integer maximizing the alignment,
\begin{equation} \label{alignansatz}
 \begin{pmatrix} Q_{11} & Q_{12} \\ Q_{21} & Q_{22} \\ Q_{31} & Q_{32} \end{pmatrix} 
 =   \begin{pmatrix}  -Q& Q+1 \\ -Q+1 & Q \\ Q+1 & Q \end{pmatrix} \, .
\end{equation}
Thus, the effective charge $Q_{\xi}$ describing the misalignment (\ref{effcharge}) can be written as
$$
Q_\xi 
 =  \frac{1}{\sqrt{2Q^2+2Q+1}}.
 $$
For $Q=10,20,30,40,$ we find $Q_{\xi}=0.067, 0.034, 0.023, 0.017$, respectively.  The resulting effective decay constants are plotted as a function of $\tau_1$ in Fig. \ref{fig:feff}.  Comparing this to \eqref{fnatural}, we see that $f_{\text{eff}}\geq 4M_{\rm Pl}$ is satisfied, and thus natural inflation is possible, for a wide range of parameters.  Furthermore, for $Q\lesssim 40$, moderate values of $N_I$ are sufficient to suppress tunneling (c.f. {\eqref{tunnelingaligned}}), so we may have further enhancement by the $N_I$ factors.

We have sufficient suppression of the third potential solving the CHC that it does not affect the inflationary dynamics. We stress again that this smallness, 
$$ 
 e^{-S_3} = e^{-2\pi ( Q_{31} \tau_1 + Q_{31} \tau_2)} \simeq e^{-2\pi (10 \cdot 10.1 + 10 \cdot 7.15)} \, ,
$$
is a consequence of the opposite signs of the charges (\ref{oppositesigns}) from the CHC with the small volumes $\tau_i$ in Table \ref{t:values}.
Finally, we predict an axion mass
\begin{equation}
	m_\xi = \frac{\Lambda_2^2}{f_{\rm eff}} \sim 10^9~ \rm{GeV}\, ,
\end{equation}
which is well above the range of dark radiation.

\begin{figure}[t]
 \begin{center}
 \includegraphics[width=14cm]{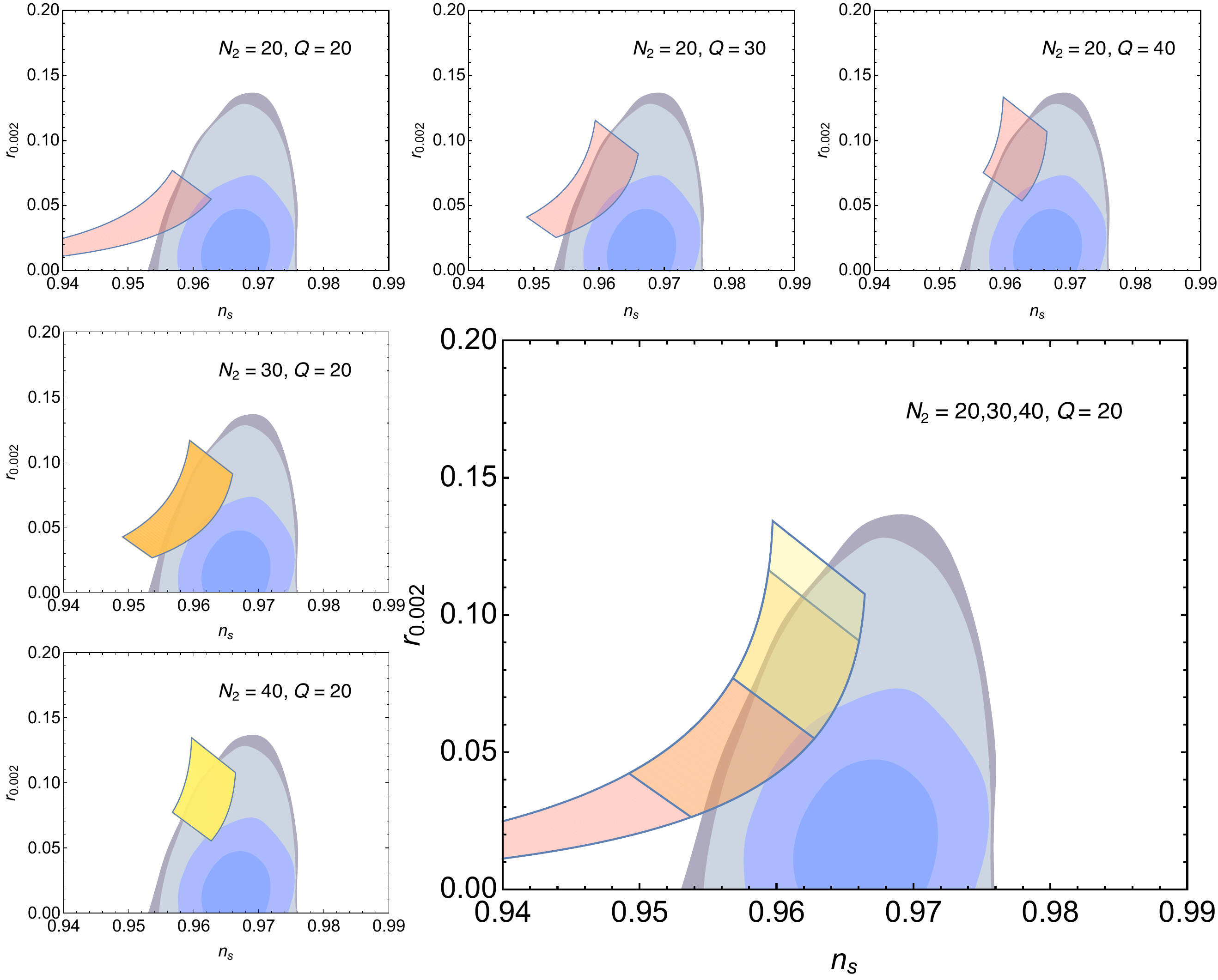}
\end{center}
\caption{Plot of the spectral index $n_s$ and the tensor-to-scalar ratio $r_{0.002}$. From top to bottom, we have $N_2=20,30,40$, respectively. From left to right, we have $Q=20,30,40$, respectively. We take $\tau_2 = \sqrt{2} \tau_1$ and every figure is drawn in the range $10 \le \tau_1 \le 20$ and $50 \le N_e \le 60$.}
\label{fig:nsrgraph}
\end{figure}

In Fig. \ref{fig:nsrgraph}, we draw predictions for the inflation parameters: the spectral index $n_s$ and the tensor-to-scalar ratio $r$ and compare them with the Planck 2018 results \cite{Aghanim:2018eyx}. We use the ansatz (\ref{alignansatz}) for the charges.  In each plot, the lengths of the shaded regions correspond to $10 \le \tau_1 \le 20$, and the widths correspond to $50 \le N_e \le 60$. 
Large values of $Q$ and $N_2$ result in enhancement of the effective decay constant. Of course, the  original decay constants of each axion are larger for smaller $\tau_i$s. Since natural inflation is described by only one parameter, the (effective) decay constant $f_{\rm eff}$, the relation between $n_s$ and $r$ in (\ref{tensortoscalar}) is fixed, and their values always lie in the same strip for different values of $\tau_i$, $Q$ and $N_2$.

\section{Discussion}

In this work, we embedded aligned natural inflation the Large Volume Scenario of type IIB string theory and extracted predictions for cosmological observables in terms of the underlying geometry.  Axions naturally emerge from compactifications of higher-dimensional supersymmetric theories and are associated with the volume moduli of four-cycles in the internal manifold. The shift symmetry of the four-form field is broken by D7-branes wrapped on these cycles and generates an axion potential, relating the instanton action and the axion decay constants. 

We have stabilized the moduli of a Calabi--Yau orientifold compactification of Type IIB string theory within the Large Volume Scenario. As a reference geometry, we considered a toric variety of a degree-12 polynomial in $\mathbb{CP}^4_{[1,1,2,2,6]}$ with an additional blow-up, as well as a degree-8 polynomial in $\mathbb{CP}^4_{[1,1,2,2,2]}.$ We also studied general predictions applicable to any number of K\"ahler moduli, following the general analysis \cite{Cicoli:2008va}. 

Generically, the internal geometry has multiple four-cycles. The LVS mechanism works in two steps. First, the small cycles and associated axions are stabilized by the interplay of perturbative and non-perturbative corrections, fixing the overall volume.  Following this, the remaining ``bulk'' moduli orthogonal to the volume modulus can then be stabilized by one-loop corrections to the K\"ahler potential that arise due to KK and winding modes. 
At this stage, the axions associated with large bulk cycles 
are unstabilized.  {However, they may obtain hierarchically small masses from the instanton contributions of D-branes wrapping the corresponding bulk cycles.   With multiple bulk cycles, provided the compactification volume is not too large, we may have multiple axions contributing to natural inflation. 
These may be aligned via the Kim--Nilles--Peloso mechanism and also further enhanced, if there are stacks of multiple D7-branes, by the large rank of the associated strongly-coupled gauge group.

The relative sizes of the decay constants, $f_i/f_j$, $i\ne j$, are fixed by the ratios of bulk moduli, which are in turn determined by string-loop corrections. The ratios are subject to a number of restrictions:
\begin{enumerate}[label=(\arabic*)]
\item The condition that the stabilized moduli lie in the K\"ahler cone;
\item The convex hull condition (CHC) of the weak gravity conjecture;
\item The relation between the power spectrum of the scalar perturbations and the height of the axion potential.
\end{enumerate}
These have interesting tensions: conditions (1) and (2) prefer the ratio $f_1/f_2$ to be large and small, respectively, and (3) favors ratios almost equal to one.

It is interesting to see that, in our example ${\mathbb C}\P^4_{[1,1,2,2,6]}(12)$, 
the above constraints, given in \eqref{nearlyisotropic}, \eqref{nearlyisotropic2} and \eqref{Pconstraint}, tend to favor nearly isotropic geometry and thus aligned axions with similar decay constants,
\begin{equation}
  \frac{f_1^2}{f_2^2} \simeq \frac{\tau_2^2}{2\tau_1^2} \sim 1 \, .
\end{equation}
We should mention that this behavior may be particular to this example because condition (1) is highly dependent on the details of the geometry. It would be interesting to see whether this preference is generic.

We have clarified the KNP alignment scenario as well as the convex hull condition by separating the charges and the axion decay constants in them and track the individual contributions from the charge lattice, the K\"ahler potential and the rank of the strongly-coupled gauge theory. We verified the normalization of the CHC, finding that is satisfied when the convex hull spanned by the charge-to-mass vectors contain the circle of radius $\r = \sqrt{2/3}.$

In principle we should be able to calculate all properties explicitly from a given manifold, once we know the geometry, brane configuration and supersymmetry breaking structure.  The biggest obstacle to realizing this is our ignorance of the details of the one-loop corrections for a general smooth manifold. Here we took the bulk cycle volumes to be free parameters and analyzed the allowed vacua.  One compelling hint is that all of the above restrictions favor specific ranges for the ratios of bulk cycles, which in the example of ${\mathbb C}\P^4_{[1,1,2,2,6]}(12)$ correspond to nearly isotropic geometry, which is consistent with general expectations from the one-loop correction. It would be interesting to analyze geometries with more bulk axions to see if this pattern is repeated.  Furthermore, if the exact form of the one-loop corrections could be obtained, it would be interesting to find out whether the predicted ratios of the moduli in a given compactification lie in the range satisfying the three conditions above.

From our explicit embedding of aligned natural inflation, we obtained cosmological observables such as the spectral index and the tensor-to-scalar ratio, finding that they are consistent with the current cosmological data. If we allow for a larger $e$-folding number, $N_{e}\sim 70$, which is also consistent with current observations, we may obtain a better fit to the central values of those parameters.  

Since the axion is heavy enough to drive inflation,
\begin{equation}
 m_a \sim  M_{\rm Pl} e^{-c \V^{2/3}} \sim 10^9 \text{ GeV} \, ,
 \end{equation}
in this sceanrio it would not contribute to dark radiation. 
Since the scale of inflation prefers a not-so-Large volume, a more careful analysis of the perturbative expansion in negative powers of $\V$ should be necessary.
There can also be a correction at the same order in $\V$ as the string loop corrections but not suppressed by $g_s$.  It arises as a tree-level ${\cal O}(\alpha^{\prime 2})$ correction to the volume \cite{Grimm:2013gma,Pedro:2013qga}, and originates from the intersection between D7-branes and O7-planes.  This has been argued to place a strong upper bound on the allowed volume for which the LVS expansion is trustable.  If present, these corrections would cast significant doubts on on the moduli stabilization scenario we have presented, and further analysis would be required to obtain cosmological predictions. 

One important issue we have not addressed in this work is the location of the Standard Model.  After inflation, energy needs to be transferred from the axion to the visible sector in order to induce reheating and initiate the hot Big Bang expansion.  Presumably this would require a large enough coupling to the bulk axion, for example via an axion-photon-photon coupling, which may lead to other testable predictions in the CMB.  It would be interesting to investigate further in this direction.  Furthermore, since we have high-scale supersymmetry breaking, an alternative resolution to the electroweak hierarchy problem would be required.  It would be interesting to see if a candidate resolution, such as the relaxion scenario \cite{Graham:2015cka}, could be integrated into our setup.

As for the origin of the brane configuration, we may gain hints from the unification picture via brane recombination, which is a natural consequence of the RR tadpole cancellation condition \cite{Choi:2005pk,Choi:2006hm,Angus:2019tct}. If supersymmetry is preserved, the brane configuration may arise via a zero-energy transition from a simple setup, such as the type I string, transitioning to various other vacua.
It would be interesting to combine this model with a brane construction of the Standard Model. Its overlap with the axion sector gives us further predictions, such as the reheating temperature.

Finally, it would be worthwhile to obtain a fully top-down description of the Large Volume Scenario.  In particular, an important consistency check would be to derive an explicit uplifting potential to lift the vacuum to Minkowski or de Sitter, and then verify whether the full top-down LVS construction is consistent and stable, along with our brane configuration and inflation model.

\subsection*{Note added} While finishing this project, a related work appeared \cite{Broeckel:2021dpz}. Although that work focuses on realizing the QCD axion, there is some overlap in the analysis.

\subsubsection*{Acknowledgments}
We are grateful to Michele Cicoli, Joe Conlon, Kwang Sik Jeong, Jihn E. Kim, Bumseok Kyae, Seung-Joo Lee, Liam McAllister, Mu-In Park, Wanil Park, Min-Seok Seo and Pablo Soler for discussions and correspondences.  SA is supported by an appointment to the JRG Program at the APCTP through the Science and Technology Promotion Fund and Lottery Fund of the Korean Government, which is also supported by the Korean Local Governments of Gyeongsangbuk-do Province and Pohang City. KSC is partly supported by the grant NRF-2018R1A2B2007163 of National Research Foundation of Korea. CSS is supported in part by the IBS under project code IBS-R018-D1.

\appendix

\section{Geometry}

In this appendix, we briefly construct an example geometry to understand the meaning of the decay constants and their constrained ranges, following \cite{Pedro:2013qga,Cicoli:2011it,Denef:2004dm}. The Calabi--Yau orientifold is given by a degree 12 polynomial
$$
 z_5^2 = P_{12} (z_1,z_2,z_3,z_4, z_6)
$$
in $\mathbb{CP}_{[1,1,2,2,6]}^4$, whose homogeneous coordinates are $z_i,i=1,2,3,4,5$. We introduce an auxiliary coordinate $z_6$ in order to embed the polynomial and assign the following scaling:
\begin{center}
\begin{tabular}{cccccc}
\hline \hline
$z_1$ & $z_2$ & $z_3$ & $z_4$ & $z_5$ & $z_6$ \\
\hline
1 & 1 & 2 & 2 & 6 & 0  \\
0 & 0 & 1 & 1 & 3 & 1 \\
\hline
\end{tabular}
\end{center}
The ordinary divisors are $\tilde D_i = \{ z_i =0 \},i=1,\dots 6$.
The columns form the vertex vectors of the toric diagram.

The homology $H_4(X,\Z)$ is spanned by the basis vectors $\tilde D_1$ and $\tilde D_4$. The K\"ahler form $J= \tilde t^1 \tilde D_1 + \tilde t^4 \tilde D_4$ spans the K\"ahler cone with non-negative coefficients, $\tilde t^1\ge0$, $\tilde t^4\ge0$. 
The nonvanishing intersection numbers are $\tilde \kappa_{144}=2$ and $\tilde \kappa_{444}= 4$. They give the volume 
\begin{equation}
\V = \tilde t^1 (\tilde t^4)^2 + \frac{2}{3} (\tilde t^4)^3 \, .
\end{equation}

The dual four-form volumes are
$$
 \tilde \tau_1 = \frac12 \int_{\tilde D_1} J^2 = (\tilde t^4)^2 \, , \qquad \tilde \tau_4=\frac12 \int_{\tilde D_4} J^2 = 2(\tilde t^1+\tilde t^4)\tilde t^4 \, .
$$ 
Inverting this gives
$$
 \V = \frac12 \sqrt{\tilde \tau_1} \left(\tilde \tau_4 -\frac{2}{3} \tilde \tau_1 \right), \quad  \tilde t^1 = \frac{\tilde \tau_4 - 2 \tilde \tau_1}{2\sqrt{\tilde \tau_1}} \, , \qquad \tilde t^4= \sqrt{\tilde \tau_1} \, .
$$
The K\"ahler cone condition is also converted as
\begin{equation}
 \tilde \tau_4 \ge 2 \tilde \tau_1 >0 \, .
\end{equation}
This guarantees the volume
to be positive, since $3\tilde \tau_4-2\tilde \tau_1 \ge 4 \tilde \tau_1 \ge 0$.
The involution is $z_5 \leftrightarrow -z_5$, defining the orientifold divisor as $D_O = \{ z_5=0 \}$. From the toric data, we have $D_O = 12 \tilde D_1 - 3 \tilde D_4.$

In the main text, we redefine 
$$ D_1 = \tilde D_1 \, , \quad 3D_2 \equiv 3 \tilde D_4 - 2 \tilde D_1 \, , $$ 
giving
$$
 \tau_1 = \tilde \tau_1\,, \qquad  \tau_2 = \tilde \tau_4 - \frac{2}{3} \tilde \tau_1\, , \qquad t^1 = \tilde t^1 + \frac23 \tilde t^2 \, , \qquad t^2 = \tilde t^2 \, .
$$
This reparametrization makes the analysis simpler because the resulting volume and the K\"ahler metric,
\begin{equation} \label{volumeapp}
 \V = t^1 (t^2)^2 = \frac12 \sqrt{\tau_1} \tau_2\, ,\qquad t^1 = \frac{\tau_2}{2 \sqrt{\tau_1}} \, , \qquad t^2 = \sqrt{\tau_1} \, ,
\end{equation}
and hence the axion decay constants are diagonal \cite{Cicoli:2011it}. The K\"ahler cone is well-defined in the original coordinates, and the boundary is $\tau_1 \ge 0, \tau_2 \ge 4 \tau_1/3$. In this basis, the only non-vanishing intersection number is $\kappa_{122}=2$, consistent with the volume \eqref{volumeapp}. A nontrivial check is $\kappa_{222} = D_2^3 = (\tilde D_4 - \frac23 \tilde D_1)^3 = \tilde D_4^3 - 2 \tilde D_4^2 \tilde D_1 = \tilde \kappa_{444} - 2 \tilde \kappa_{144} = 0$. In the new coordinate, the orientifold plane is $D_O = 10 D_1 -3 D_2$.

\end{document}